\documentclass{emulateapj}

\shorttitle{ICL}
\shortauthors{Krick, Bernstein}

\begin{document}
\newcommand\msun{\hbox{M$_{\odot}$}}
\newcommand\lsun{\hbox{L$_{\odot}$}}
\newcommand\magarc{mag arcsec$^{-2}$}

\bibliographystyle{myapj}

\title{\bf Diffuse Optical Light in Galaxy Clusters I: Abell 3888} 

\author{J.E. Krick and R.A. Bernstein}
\affil{Astronomy Department, University of Michigan, Ann Arbor, MI 48109}
\email{jkrick@umich.edu, rabernst@umich.edu}

\and

\author{K.A. Pimbblet}
\affil{Department of Physics, University of Queensland, Brisbane,
QLD 4072, Australia}
\email{pimbblet@physics.uq.edu.au}

\begin{abstract} 

We are undertaking a program to measure the characteristics of the
intracluster light (total flux, profile, color, and substructure) in a
sample of 10 galaxy clusters with a range of cluster mass, morphology,
and redshift.  We present here the methods and results for the first
cluster in that sample, A3888.  We have identified an intracluster
light (ICL) component in A3888 in $V$ and $r$ that contains $13\pm5$\%
of the total cluster light and extends to 700h$_{70}^{-1}$kpc ($\sim
0.3 r_{200}$) from the center of the cluster.  The ICL color in our
smallest radial bin is $V-r = 0.3 \pm 0.1$, similar to the central
cluster ellipticals.  The ICL is redder than the galaxies at $400 < r
< 700$h$_{70}^{-1}$kpc although the uncertainty in any one radial bin
is high.  Based on a comparison of $V-r$ color with simple stellar
models, the ICL contains a component which formed more than 7 Gyr ago
(at $z > 1$), coupled with a high metallicity ($1.0Z_{\odot} < Z_{ICL}
\la 2.5Z_{\odot}$), and a more centralized component which contains
stars formed within the past 5 Gyr (at $z \sim 1$). The profile of the
ICL can be roughly fit by a shallow exponential in the outer regions
and a steeper exponential in the central region.  We also find a
concentration of diffuse light around a small group of galaxies
1.4h$_{70}^{-1}$Mpc from the center of the cluster.  In addition, we
find 3 low surface brightness features near the cluster center which
are blue ($V-r = 0.0$) and contain a total flux of 0.1$M^*$. Based on
these observations and X-ray and galaxy morphology, we suggest that
this cluster is entering a phase of significant merging of galaxy
groups in the core, whereupon we expect the ICL fraction to grow
significantly with the formation of a cD galaxy as well as the in-fall
of groups.

\end{abstract}


\section{Introduction} 

Galaxy clusters contain a population of stars which are not members of
individual galaxies but which are bound to the cluster potential,
producing diffuse intracluster light (ICL).  This ICL component has
been found in a number of clusters through surface brightness
measurements and direct detections of resolved stars including
planetary nebulae, red giants, supernovae, novae, and globular cluster
systems.  These investigations indicate that the optical ICL comprises
between 5\% and 50\% of the total optical cluster luminosity
\citep[see][and references
therein]{feldmeier2004,gonzalez2005,zibetti2005}.  Conclusions on the
color of the ICL vary widely from blue to red, with and without color
gradients \citep[][]{schombert1988, mackie1992, gonzalez2000,
zibetti2005}.  Current measurements of the shape of the ICL generally
favor a double deVaucouleurs profile such that one function fits the
BCG and the second function fits the extended envelope
\citep{gonzalez2000,bernstein1995,zibetti2005}.  Examples of tidal
features such as plumes and bridges are found in multiple clusters as
evidence of interactions which add stars to the ICL
\citep{gregg1998,calcaneo2000,feldmeier2004}.  Long-slit spectroscopy
of A2199 shows that the intracluster stars there have the same
velocity dispersion as the cluster galaxies \citep{kelson2002},
confirming that, in at least one cluster, the intracluster stars are
not bound to the individual galaxies but trace the overall cluster
potential.  Conversely, intracluster planetary nebulae studies show
evidence for less relaxed velocities
\citep{arnaboldi2004,gerhard2005}.  There is no consensus on the
velocity distribution of intracluster stars.

The intracluster light is a fossil remnant of cluster formation and
evolution and can be used to study the dominant physical processes
involved in galaxy evolution in clusters.  Hierarchical dark matter
simulations suggest that galaxies falling into dense regions would
loose most of their mass.  When mechanisms such as radiative cooling
and star formation are included in the simulations, galaxies which are
composed of a central dense core of stars do retain most of their
stellar mass throughout cluster in-fall, but lose some stars to the
cluster potential.  State of the art simulations are able to predict
the existence of this intracluster population, but basic questions as
to its properties can only be answered by understanding which physical
mechanisms are important.  This work seeks to answer when and how
intracluster stars are formed by studying the total flux, profile
shape, color, and substructure in the ICL as a function of cluster
mass, morphology, and redshift.

Observations of the total flux in the ICL over a sample of clusters
will allow us to identify the effects of cluster environment on galaxy
evolution.  For example, a high mass cluster should have a higher ICL
fraction than low-mass clusters if ram pressure stripping or
harassment are dominant mechanisms.  In fact, simulations by both
\citet{lin2004} and \citet{murante2004} predict a strong relation
between ICL fraction and mass.  If, however, galaxy-galaxy merging is
the dominant mechanism, and most of the galaxy evolution happens early
on in cluster collapse, then the ICL should not correlate directly
with current cluster mass.  The existence of a cD galaxy in a cluster is
evidence of a rich merger history and therefore morphology should also
correlate strongly with ICL fraction.  The ICL fraction will be
affected by redshift, since with time comes an increased number of
interactions.

Observations of the color and substructure of the ICL will help to
identify the origin, formation epoch, metallicity, and possibly
progenitor morphologies of cluster galaxies.  For example, if the ICL
is as red or redder than the bright cluster ellipticals, it is likely
to be a remnant from the early epochs of cluster formation with little
recent accretion of tidally disrupted systems.  If the ICL is bluer
than the galaxies, then some recent accretion has occurred, either
from ellipticals with low metallicity or spirals with younger stellar
populations.  While multiple mechanisms are likely to play a role in
the complicated process of formation and evolution of clusters,
important constraints can come from ICL measurement in clusters with a
wide range of properties.

In addition to constraining galaxy evolution, the ICL is an important
baryonic component in clusters.  The ICL, which is typically not
included in the baryon census, will contribute to the baryon fraction
of clusters and that contribution is likely to change with time. If
the ICL fraction does indeed evolve with redshift and is a significant
fraction of the total cluster light, it will systematically bias the
inferred redshift dependence of the baryon fraction.  Recent work by
\citet{allen2004} has used a change in baryon fraction with redshift
of only a few percent to constrain cosmological parameters.  When
doing such precision cosmology it will be necessary to include ICL in
the cluster light budget.

The ICL may also play an important role in the global properties of
the intracluster medium (ICM).  It has recently been suggested that an
intracluster stellar population (ICSP) can account for at least some
amount of heating and metal enrichment of the intracluster medium
(ICM) \citep{zaritsky2004,lin2004,domainko2004}.  Considering only
supernovae (SNe) {\it within} galaxies, the full metallicity of the
ICM can not easily be accounted for \citep{lin2004}.  However, since
intracluster supernovae are {\it in situ} in the ICM, they contribute
directly to the metallicity of the ICM, and will have therefore a
direct impact on its abundance.  Although these authors find that the
ICL cannot fully account for the high abundance of the ICM ($\sim
0.3Z_{\odot}$), further studies are warranted to quantify just how
many ICSNe there are.  Even if the ICSP can not account for the full
metallicity of the ICM, it is possible that this population is
responsible for the metallicity gradient found in clusters. If true, a
correlation between ICL flux and abundance gradients in clusters
should exist.

In this paper we present the methods of this survey as well as 
measurements of the color, total flux, and profile shape for the first
cluster in our sample, A3888.  In \S \ref{sample} we discuss
characteristics of the entire sample.  Details of the observations and
reduction are presented in \S \ref{observations} and \S
\ref{reduction} including flat-fielding and sky background subtraction
methods.  In \S \ref{analysis} we discuss object detection and removal
as well as cluster membership.  In \S \ref{results} we describe
results followed by a discussion of accuracy limits in \S \ref{noise}.
In \S \ref{discuss} we present a discussion of the results.
Conclusions are summarized in \S \ref{conclude}.

Throughout this paper we use $H_0=70$km/s/Mpc, $\Omega_M$ = 0.3,
$\Omega_\Lambda$ = 0.7 which gives 3.5 kpc/arcsecond at the distance
of A3888.


\section{The Sample}
\label{sample}

The ten galaxy clusters in our survey were selected to meet two
general criteria. First, each cluster has a published X-ray luminosity which
guarantees the presence of a cluster and provides an estimate of the
cluster's mass. Second, all are at high galactic and ecliptic latitude
along lines of sight with low HI column density.  This minimizes
complications due to scattered light from galactic stars and zodiacal
dust and from variable extinction across the cluster field.  Of the
clusters that meet the above qualifications, we selected ten clusters
as the beginning of a statistical sample which is representative of a
wide range in cluster characteristics, namely redshift, morphology,
spatial projected density (richness), and X-ray luminosity (mass).

To the extent possible, we also selected clusters for which mass
estimates and membership information are available in the literature.
For example, in addition to published X-ray masses, three of the
clusters have mass estimates from gravitational lensing
measurements. Published redshift surveys provide velocity dispersions
and membership information for all but 2 clusters in the survey. Those
2 clusters, as well as 5 others with small numbers of published
velocities, were included in a redshift survey we undertook with IMACS
on Magellan I (Baade). With these additional observations the physical
properties of all clusters in our sample can be compared to the ICL
characteristics.  Table 1 lists the relevant
information for the cluster sample.

The sample is divided into a ``low'' ($0.05<z<0.1$) and ``high''
($0.15 < z< 0.3$) redshift range which we have observed with the 1
meter Swope and 2.5 meter Du Pont telescope respectively.  The bottom
end of the redshift range is limited by the field of view of the 1
meter telescope and detector , which corresponds to $0.9\times1.4
h_{70}^{-1}$ Mpc ($14.8\arcmin\times22.8\arcmin$) at z = 0.05.  This
field of view allows us to measure the ICL as well as off-cluster
background flux in the same image for all clusters in the sample.  The
top end of the redshift range reflects X-ray data availability and the
increasing difficulty of measuring diffuse sources at high redshift
due to $(1+z)^4$ surface brightness dimming.

\subsection{A3888}
\label{A3888}

This paper focuses on one cluster in our sample, A3888, which is a
richness class 2 cluster at z=0.15 \citep{abell1989}.  This
Bautz-Morgan type I-II cluster has no cD galaxy; instead the core is
comprised of 3 distinct sub-clumps of multiple galaxies each. At least
2 galaxies in each of these clumps are confirmed members
\citep{teague1990,pimbblet2002}.  On large scales (286, 535, and 714
$h^{-1}_{70}$kpc) \citet{Girardi1997} find a unimodal distribution for
this cluster with no detected substructures in either the galaxy
spatial or velocity distribution.  The projected spatial distribution
of galaxies in A3888 is slightly elongated with an ellipticity of 0.43
\citep{struble1994}.  X-ray surface brightness from XMM observations
also indicate an elongated, single-peaked distribution for the hot
gas.  The cluster contains an X-ray bright Seyfert I galaxy located at
a projected distance of roughly $600 h_{70}^{-1}$kpc from the cluster
center \citep{reimers1996}.

The mass of A3888 can be estimated from two different sets of
observations.  \citet{reiprich2002} calculate gravitational mass based
on pointed ROSAT PSPC count rates and the ROSAT-ASCA Lx-Tx relation
\citep{markevitch1998}.  Assuming an isothermal distribution and
employing hydrostatic equilibrium, they find $M_{200}$ = 25.5$\pm
^{10.8} _{7.4}\times10^{14} h_{70}^{-1}$ \msun, where $r_{200}$ =
2.8$h^{-1}_{70}$Mpc which is defined as the radius within which the
mean mass density is equal to two hundred times the critical density.
In a complementary method, mass can be determined from published
galaxy velocity dispersions.  Based on redshifts for 50 member
galaxies located within a radius of 3.11$h_{70}^{-1}$ Mpc
\citep{teague1990} and using the method described by
\citet{girardi2001}, we find that the mass of A3888 within $r_{200}$
is $M_{200}$ = 40.2$\pm ^{10.6} _{7.4}\times10^{14} h_{70}^{-1}$
\msun.  For the purpose of this work, these two mass estimates are in
good agreement, particularly since this cluster is elongated and
likely not in dynamic equilibrium.


\section{Observations}
\label{observations}

Observations for the entire sample of 10 clusters have been completed.
The ``high'' redshift observations were made with the du Pont 2.5m
telescope at Las Campanas Observatory.  We used the thinned,
$2048\times2048$ Tektronix ``Tek\#5'' CCD with a $3 e^-/$count gain
and $7e^-$ readnoise.  The pixel scale is 0.259\arcsec/pixel ($15\mu$
pixels), so that the full field of view is 8.8\arcmin\ on a side,
corresponding to 1.8$h_{70}^{-1}$Mpc per frame.  Data was taken in two
filters, gunn-$r$ ($\lambda_0 = 6550$ \AA) and $V$ ($\lambda_0 = 5400$
\AA). These filters were selected to provide some color constraint on
the stellar populations in the ICL, while avoiding flat-fielding
difficulties at longer wavelengths and prohibitive sky brightness at
shorter wavelengths.

Observing runs occurred on August 19-25, 1998, September 2-10, 1999,
and September 22-27, 2000.  Specifically, A3888 was observed on the
nights of September 2 and 8, 1999 and September 22-25, 2000. Both
observing runs took place in the days leading up to new moon.
September 2, 1999 was the only non-photometric night, and only 3
cluster images were taken on that night.  These were individually tied
to the photometric data.  The average seeing during the 1999 and 2000
run was 1.77 and 0.93 arcseconds respectively.  Across both runs, we
observed A3888 for an average of 5 hours in each band. In addition to
the cluster frames, night sky flats were obtained in nearby,
off-cluster, ``blank'' regions of the sky with total exposure times
roughly equal to one third of the integration times on cluster
targets.  Night sky flats were taken in all moon conditions.  Typical
$V-$ and $r-$band sky levels during the run were $21.3$ and $21.1$
\magarc, respectively.

Cluster images were dithered by one third of the field of view between
exposures.  The large overlap from the dithering pattern gives us
ample area for linking background values from the neighboring cluster
images.  Observing the cluster in multiple positions on the chip is
beneficial because upon combination large-scale flat-fielding
fluctuations will be reduced.  Integration times were typically 900
seconds as a compromise between signal-to-noise and moderating the
number of saturated stars.

Observations of the ``low'' redshift clusters will be discussed in a
future paper.

\section{Reduction}
\label{reduction}

In order to create a single, mosaiced image of the cluster with a
uniform background level and accurate resolved--source fluxes, the
images must be bias and dark subtracted, flat--fielded, flux
calibrated, background--subtracted, extinction corrected, and
registered before combining. These issues are dealt with as described
below.  


\subsection{Bias and Dark Subtraction}

Pre-processing of the data, including overscan, bias, and dark
subtraction, was done in the standard manner using mainly IRAF tasks.
The average bias level was stable at $\sim 800$ counts, changing by
1\% throughout the night.  There is structure in the bias in the form
of random fluctuations, as well as a highly-repeatable, large-scale
ramping in the first 500 pixels of every row.  To remove this
structure, we first fit an 8th order polynomial to 140 overscan
columns and subtract that fit, column by column, from each image.  We
further average together ten bias frames per night with $3 \sigma$
cosmic ray rejection and then boxcar smooth in the vertical direction
before subtracting from the data.  We choose to smooth in the vertical
direction because we have already removed vertical structure in the
previous processing step.  Test-reduction of the bias frames
themselves with this procedure reveals no remaining visible structure
and each frame has a mean level of 0 counts to within $\pm 0.05$
counts.

Twenty-five dark exposures were taken per observing run.  We averaged
these together with $3 \sigma$ rejection to look for structure or
significant count levels in the dark current.  The mean dark count is
0.6 counts/900s, which is less than $0.08\%$ of the sky level, and is
therefore not significant.  However, even at this small count level,
there is some vertical structure in the dark which amounts to 1
count/900s over the whole image.  To remove this large-scale structure
from the data images, the combined dark frame was median smoothed over
$9\times9$ pixels ($2.3\arcsec$), scaled by the exposure time, and
subtracted from the program frames.  Small scale variations were not
present in the dark.  Errors in both the bias and dark subtraction due
to structure in the residuals are an additive offset to the background
level.  These are included in our final error budget based on an
empirical measurement of the stability of the background level in the
final combined image (see \S\ref{noise}).


\subsection{Flat Fielding}

The accuracy of any low surface brightness measurement is limited by
fluctuations on the background level.  A major contributor to those
fluctuations is the the large-scale flat-fielding accuracy.
Pixel--to--pixel sensitivity variations were corrected in all cluster
and night-sky flat images using nightly, high S/N, median-combined
dome flats with 70,000 -- 90,000 total counts.  After this step, a
large-scale illumination pattern of order 1\% remains across the chip.
This was removed using combined night-sky flats of ``blank'' regions
of the sky.  To make these night-sky flats, objects in the individual
blank sky frames were first masked before combination.  We used
SExtractor \citep{bertin1996} to identify all sources with a minimum
of 6 pixels and a total flux of $2 \sigma$ above the sky background.
Mask sizes were increased by 4-7 pixels over the semi-major and
semi-minor axes from the object catalogs to insure object rejection.
The masked images were then median combined with $2 \sigma$ rejection.
This produced an image with no evident residual flux from sources and
kept the large scale illumination pattern intact.  Fluctuations are
less than 0.1\% peak to peak on $10\arcsec$ scales.  The final
combined night-sky flats were then median smoothed $7\times7$ pixels
($2\arcsec$), normalized, and divided into the program images.  The
illumination pattern was stable among images taken during the same
moon phase.  Program images were corrected only with night sky flats
taken in conditions of similar moon.  The contribution of flat
fielding to our final error budget is measured empirically, as
described in \S\ref{noise}.


\subsection{Non-linearity}

Although the ICL measurement is based on a low number of counts,
photometric calibration is based on bright standard stars.  Accurate
calibration is then dependent upon the CCD having a linear response to
flux.  To ascertain if Tek\#5 was linear with flux over a wide dynamic
range, a consecutive chain of dome-flat images were taken, with
exposure times of 2 - 100 seconds, corresponding to approximately $300
- 15,000$ counts per pixel.  Multiple passes through the exposure time
sequence (increasing and decreasing) were made to rule out any effects
from fluctuating lamp flux.  We find that the Tek\#5 CCD does have an
approximately $2\%$ non-linearity, which we fit with a second order
polynomial and corrected for in all the data.  The same functional fit
was found for both the 1999 and 2000 data. Note that exposure times
used for all observations are long enough that shutter performance is
not a problem.  The uncertainty in the linearity correction is
incorporated in the total photometric uncertainty discussed below.


\subsection{Photometric Calibration}
\label{photometry}

Photometric calibration was performed in the usual manner. Fifty to
seventy standard stars \citep{landolt1992, jorgensen1994} were
observed per night per filter over a range of airmasses.  Stellar
magnitudes were measured with an aperture size of $5\times$ the full
width at half maximum ({\sc fwhm}), where the {\sc fwhm} of the stars
in the images was determined using SExtractor. We choose this aperture
size as a compromise between aperture correction and added background
noise. Photometric nights were analyzed together; solutions were found
in each filter for a zero-point and extinction coefficient with an
{\sc RMS} of 0.03 magnitudes ($r$) in Sept. 1999 and 0.02 magnitudes
($r$ and $V$) in Sept. 2000.  These uncertainties are a small
contribution to our final error budget, but are included for
completeness as discussed in \S\ref{noise}.  Observing the same
cluster field for long periods throughout the night allows us to
measure an extinction coefficient from stars in the cluster fields,
which we find is fully consistent with the extinction coefficient
measured from the standard stars.  The three exposures taken in
non-photometric conditions were individually tied to the photometric
data using roughly 10 stars well distributed around each frame to find
the effective extinction for that frame.  We find a standard deviation
of 0.03 within each frame, with no spatial gradient in the residuals.

We have compared our $V-$ and $r-$band magnitudes for hundreds of
galaxies in the cluster with $R-$band magnitudes from the Las
Campanas/AAT Rich Cluster Survey \citep[LARCS,][]{pimbblet2002}.  To
the detection limit of the LARCS photometry, and adopting a single
average galaxy color to convert between filters, the two samples are
consistent with an {\sc RMS} scatter of 0.07 magnitudes.


\subsection{Sky Background Subtraction}

An important issue for accurate surface brightness measurement is
accurate identification of the background sky level.  The off-cluster
background level in any image is a combination of atmospheric emission
(airglow) and light from extra-terrestrial sources (zodiacal light,
moonlight, starlight, starlight scattered off of galactic
dust). Zodiacal light comes from solar photons scattered off of
ecliptic dust and is therefore concentrated in the ecliptic plane,
which, along with the galactic plane we were careful to avoid in
sample selection, so the extra-terrestrial background light will not
vary spatially.  Light from the extra-terrestrial sources will
additionally be scattered into the field of view by the atmosphere.
Airglow is emission from the recombination of electrons in the earth's
atmosphere which were excited during the day by solar photons, and as
such is a function of solar activity, time elapsed since sunset, and
geomagnetic latitude \citep{leinert1998}.  Airglow and atmospheric
scattering vary throughout the night, moonlight varies from night to
night, and zodiacal light varies from year to year.  The background
values from frame to frame correspondingly vary temporally by up to
10\% throughout one run and 20\% from year to year.

Due to the temporal variations in the background, it is necessary to
link the off-cluster backgrounds from adjacent frames to create one
single background of zero counts for the entire cluster mosaic before
averaging together frames. To determine the background on each
individual frame we measure average counts in approximately twenty
$20\times20$ pixel regions across the frame.  Regions are chosen
individually by hand to be a representative sample of all areas of the
frame that are more distant than $0.8 h_{70}^{-1}$Mpc from the center
of the cluster.  This is well beyond the radius at which ICL
components have been identified in other clusters
\citep{feldmeier2002, gonzalez2005}, and is also beyond the radius at
which we detect any diffuse light in A3888.  The average of these
background regions for each frame is subtracted from the data,
bringing every frame to a zero background.  The accuracy of the
background subtraction will be discussed in \S \ref{noise}.


\subsection{Extinction Correction}

After background subtraction, all flux in the frame originates above
the atmosphere, and is subject to atmospheric extinction (large angle
scattering out of the line of sight). This is equally true of resolved
sources and diffuse sources less than several degrees in extent.
While extinction corrections are usually applied to individual
resolved sources, that is not possible with the diffuse ICL.  We
correct entire cluster images for this extinction by multiplying each
individual image by $10^{\tau\chi/2.5}$, where $\chi$ is the airmass
and $\tau$ is the fitted extinction term from the photometric
solution.  This multiplicative correction is between 1.06 and 1.29 for
the airmass range of our A3888 observations.



\subsection{Registration \& Distortion}

To combine images, we align all 41 individual frames to one central
reference frame.  SExtractor positions of approximately 10 stars in
each frame are used as input coordinates to the IRAF tasks {\sc
geomap} and {\sc geotran}, which find and apply x and y shifts and
rotations between images.  The {\sc geotran} solution is accurate to
0.01 pixels (RMS).  As an independent check of registration accuracy,
we confirm that the center coordinates of stars in the original
images, as compared to the combined image, are the same to within 0.01
pixels.  This uncertainty is negligible for our measurement which is
made on much larger scales.  In addition, the ellipticities of
individual stars do not change with image combination, suggesting that
no systematic errors in registration exist.  Stellar ellipticities
also show no variation across the frame, suggesting that there are no
significant image distortions.


\subsection{Image Combination}
\label{combination}

\begin{figure}[h]
\includegraphics[width=3.5in]{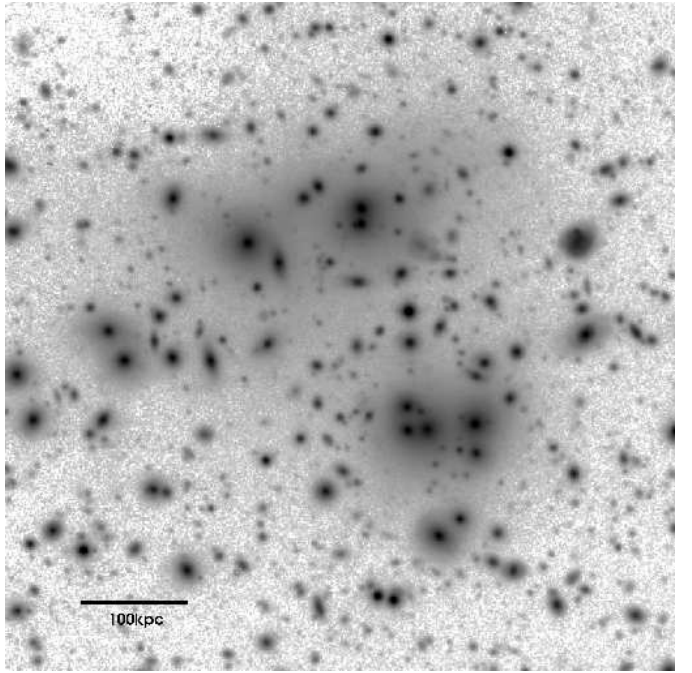}
\caption
  {The central $\sim1$Mpc (4.9 arcmin) of A3888 
  shown in the $V$ band. The gray--scale is linear 
  over the range 20.4--29.5 mag arcsec$^{-2}$.  
  Note the three main groups of galaxies near the
  center of the cluster. A dozen objects in this 
  image are stars; the rest are galaxies.}
\label{fig:centerVmu}
\end{figure}

After pre-processing, background subtraction, extinction correction,
and registration, we combined the images using the IRAF routine {\sc
imcombine} with a rejection of $-3.5\sigma$ and $+4.5\sigma$.  This
range was chosen as a compromise between rejecting the cosmic rays
(CRs) and allowing for some seeing variations in the peak flux of
stars.  In total, 16 and 25 900s exposures in the $V-$ and $r-$ bands,
respectively, were averaged together.  The final combined image is
4096 pixels ($3.6h_{70}^{-1}$Mpc) on a side. The central region
(approximately $1 h_{70}^{-1}$Mpc on a side) of the final combined
$V-$band image is shown in Figure \ref{fig:centerVmu}.

\section{Analysis}
\label{analysis}

\subsection{Object Detection}
\label{detection}

We use SExtractor both to find all objects in the combined frames and
to determine their shape parameters.  The detection threshold in both
the $V$ and $r$ images was defined such that objects have a minimum of 6
contiguous pixels, each of which are greater than $1.5\sigma$ above
the background sky level.  This corresponds to a minimum surface
brightness of 26.0 \magarc\ in $V$ and 26.4 \magarc\ in $r$.  The faintest
object in the catalog has a total magnitude of 27.0 mag in $V$ and 27.4
mag in $r$, however we are complete only to 24.8 mag in $V$ and 24.5 mag
in $r$.  We choose these parameters as a compromise between detecting
faint objects in high signal-to-noise regions and rejecting noise
fluctuations in low signal-to-noise regions.  Shape parameters are
determined in SExtractor using only those pixels above the detection
threshold.


\subsection{Object Removal \& Masking}
\label{remove_obj}

To measure the ICL we remove all detected objects from the frame by
either subtraction of an analytical profile or masking. Details of
this process are described below.


\subsubsection{Stars}
\label{star}
\begin{figure}
\includegraphics[width=3.5in]{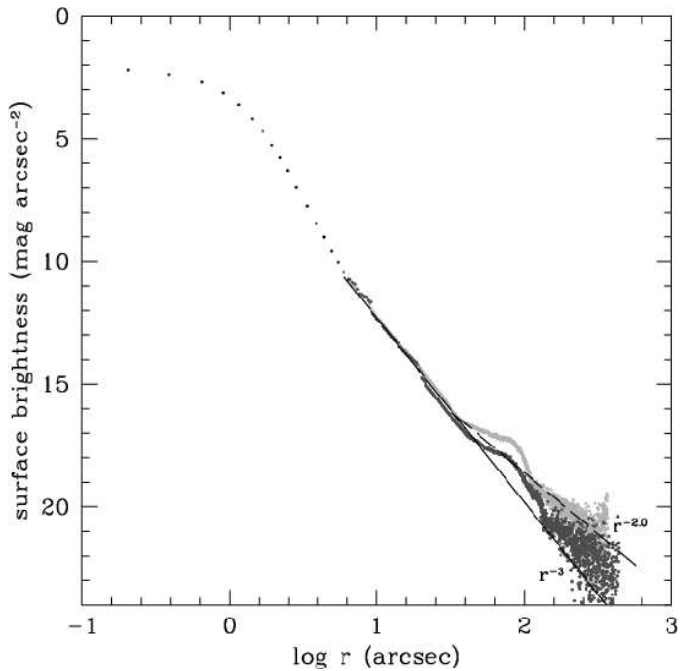}
\caption[psf]{The PSF of the 100-inch du Pont telescope at Las
  Campanas Observatory.  The y-axis shows surface brightness scaled to
  correspond to the total flux of a zero magnitude star.  The profile
  within 5\arcsec\ was measured from unsaturated stars and can be
  affected by seeing. The outer profile was measured from two stars
  with super-saturated cores imaged in two different positions on the
  CCD, on two different observing runs.  The bump in the profile at
  100\arcsec\ is likely related to the position of the star in the
  focal plane.  If the core of a star is imaged off the CCD, its
  profile does not show this feature, suggesting that the feature is
  caused by a reflection off the CCD itself.  The outer surface
  brightness profile decreases as roughly $r^{-3}$, shown by the solid
  line.  An $r^{-2.0}$ profile is plotted to show the range in
  slopes.}
\label{fig:psf}
\end{figure}

Scattered light in the telescope and atmosphere produce an extended
point spread function (PSF) for all objects.  To correct for this
effect, we determine the telescope PSF using the profiles of a
collection of stars from super-saturated 4th mag stars to unsaturated
14th magnitude stars.  The radial profiles of these stars were fit
together to form one PSF such that the extremely saturated star was
used to create the profile at large radii and the unsaturated stars
were used for the inner portion of the profile.  This allows us to
create an accurate PSF to a radius of $7$\arcmin, shown in Figure
\ref{fig:psf}.

The inner region of the PSF is well fit by a Moffat function.  The
outer region is well fit by $r^{-3}$.  There is a small additional
halo of light at roughly 50 - 100\arcsec (200-400pix) around stars
imaged on the CCD.  Images of saturated stars located off of the field
of view of the detector do not show this halo, indicating that it is
due to a reflection of light off of the CCD itself. We find that
roughly 1\% of the total flux in the star is in this halo. There are
13 saturated stars within 3.8 Mpc of A3888 ranging from 11.6 - 15.2
$V$ magnitudes.  The nearest three saturated stars are 0.6, 0.8, and
1.0 $h_{70}^{-1}$Mpc from the cluster center and have 14.6, 13.4, and
11.6 $V$ magnitudes, respectively.  These stars do not directly
contribute to the ICL measurement because they are not near enough to
the center, do not have very bright magnitudes, and the PSF does not
put very much power into the wings.  We do a careful job of background
subtraction, by tying to off-cluster flux, so that the PSF also does
not affect the background measurement.

For each individual, non-saturated star, we subtract a scaled $r^{-3}$
profile from the frame in addition to masking the inner
$30$\arcsec\ of the profile (the region which follows a Moffat
profile).  Since we do not have accurate magnitudes for the saturated
stars in our own data, and to be as cautious as possible with the PSF
wings, we have assumed the brightest possible magnitudes for these
stars given the full USNO catalog errors.  We then subtract a stellar
profile with that magnitude and produce a large mask to cover the
inner regions and any bleeding.  We can afford to be liberal with
our saturated star masking since there are very few saturated stars
and none of them are near the center of the cluster where we need the
unmasked area to measure any possible ICL.


\subsubsection{Galaxies}
\label{galaxies}

To make an ICL measurement we would ideally like to subtract a scaled
analytical profile for each galaxy that would leave no residuals and
would allow us to recover the area on the sky covered by cluster
galaxies.  We have attempted to do this using 3 publicly available
algorithms: GIM2D \citep{Simard2002}, Galfit \citep{peng2002}, and the
IRAF task {\sc ellipse} \citep{Jedrzejewski1987}. With these
algorithms, we have employed a wide range of surface brightness
profiles, including deVaucouleurs, Sersic, exponential profiles, and
combinations thereof.  In addition, we have used iterative techniques
to alternately fit and remove galaxies in crowded fields.  The
technical challenges in fitting the galaxies, including galaxy
deblending, PSF effects and deconvolution, 2D profile fitting, and
speed in performing many Fourier transforms have been previously
discussed by several groups \citep[see for example][for a
review]{peng2002}.

\begin{figure}
\epsscale{.3}
\plotone{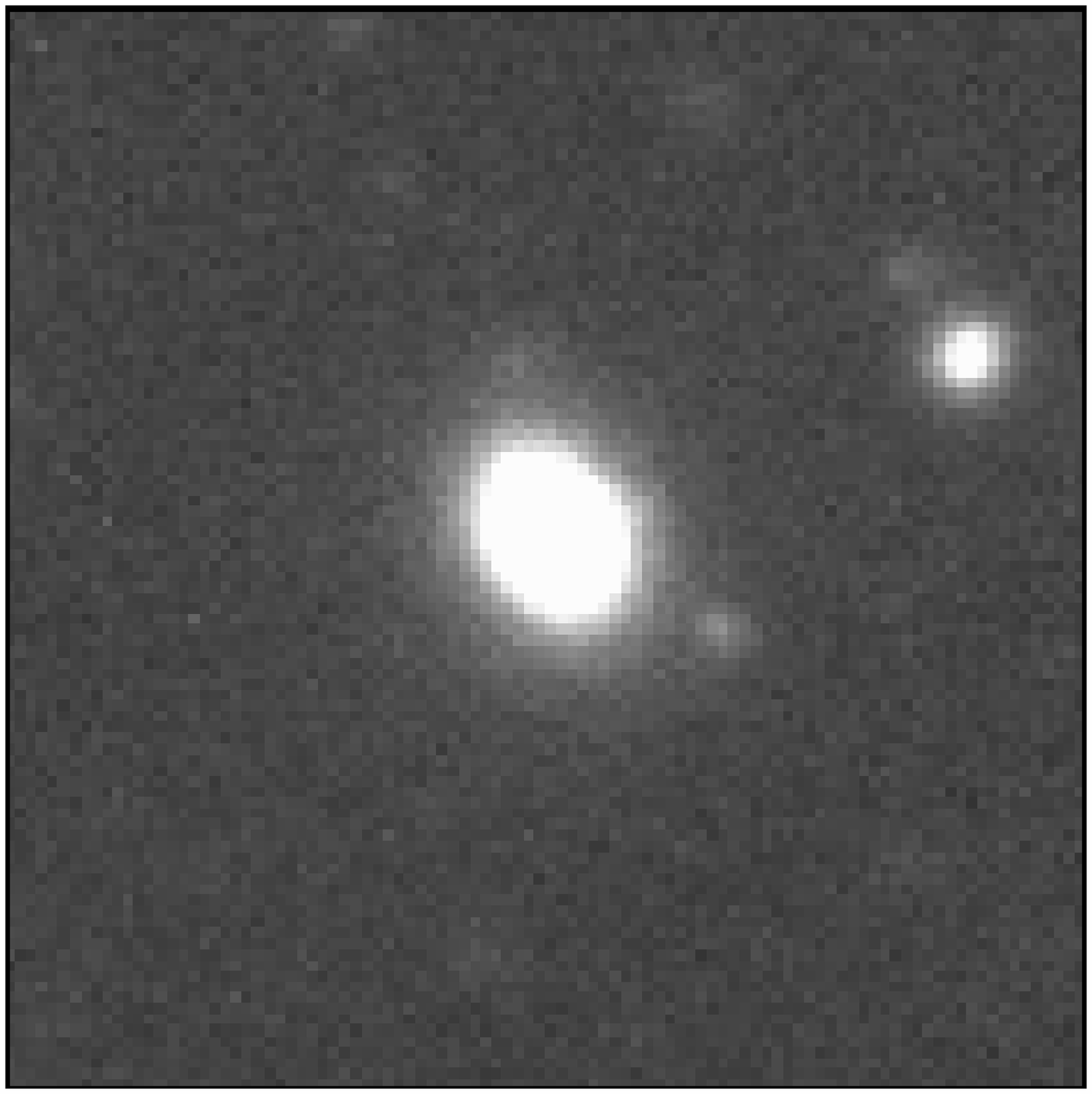}
\plotone{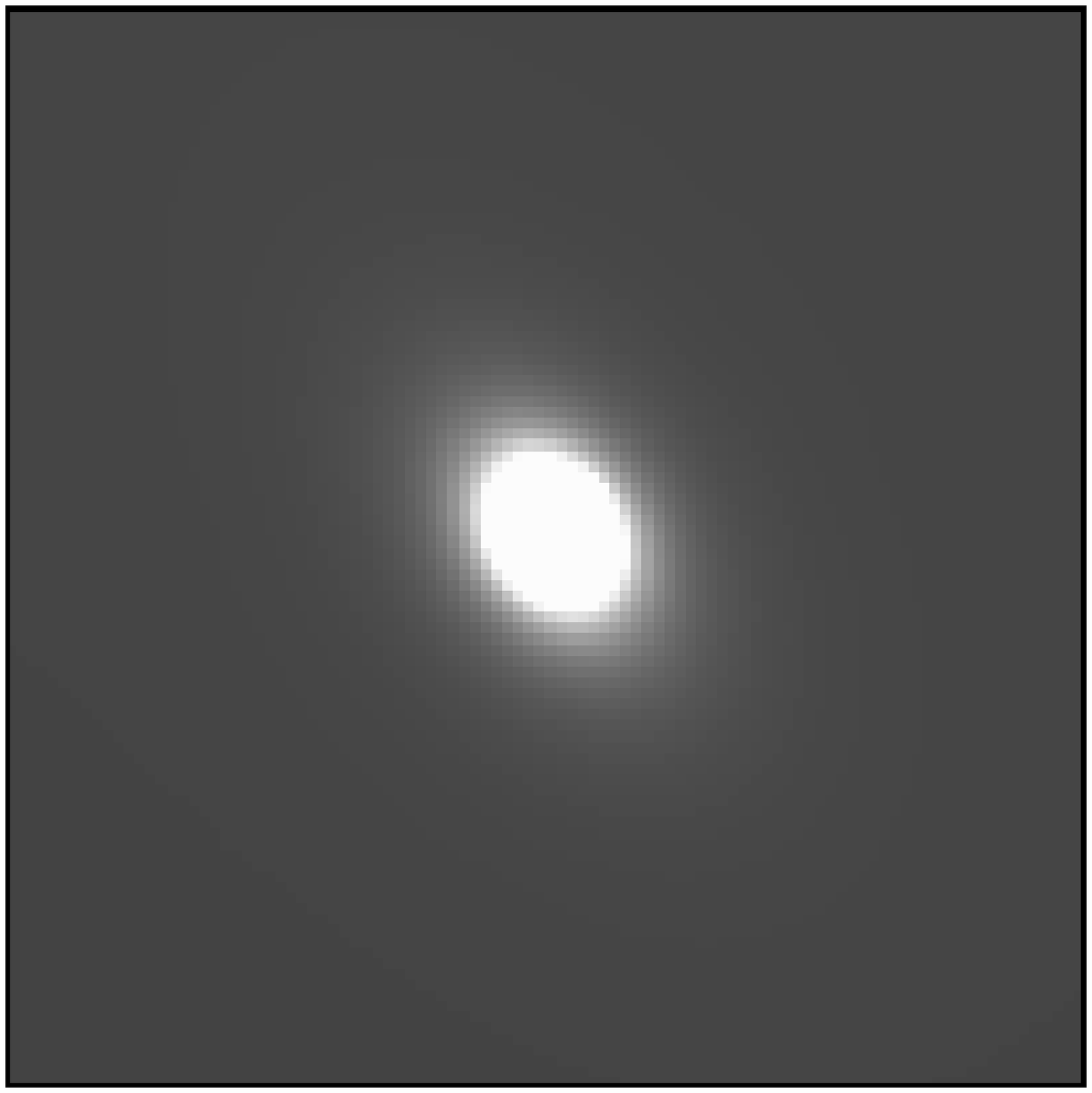}
\plotone{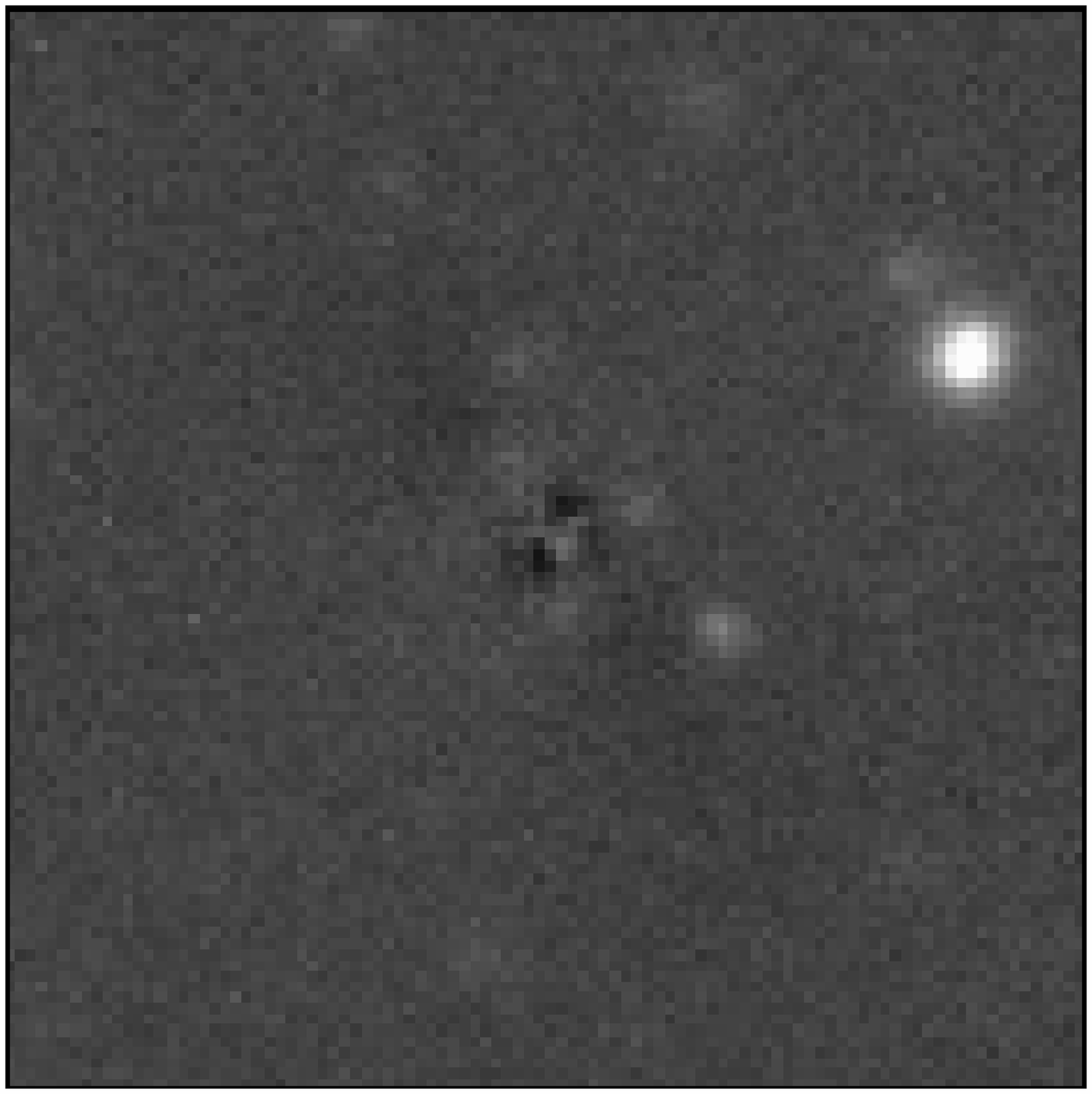}
\plotone{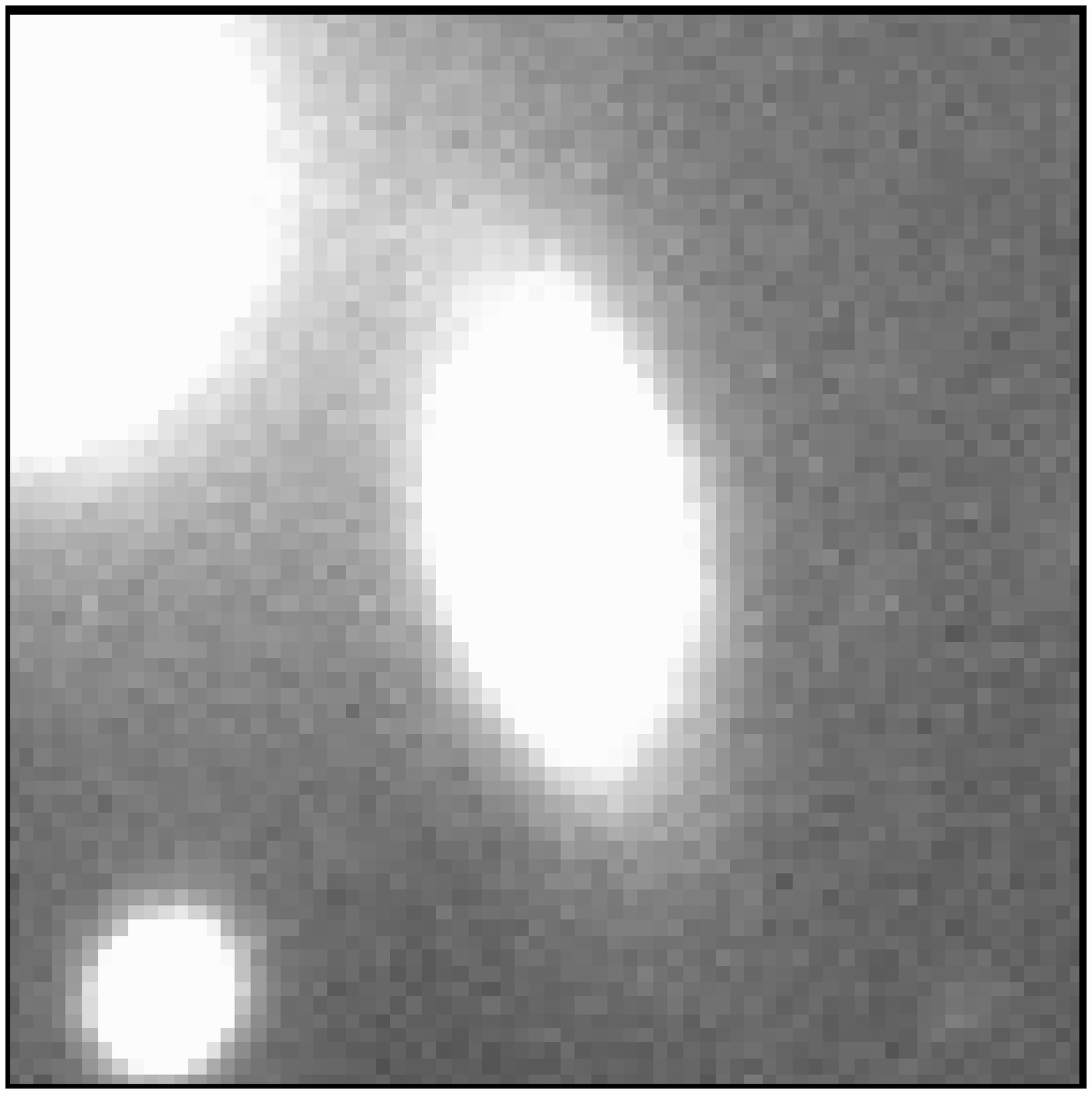}
\plotone{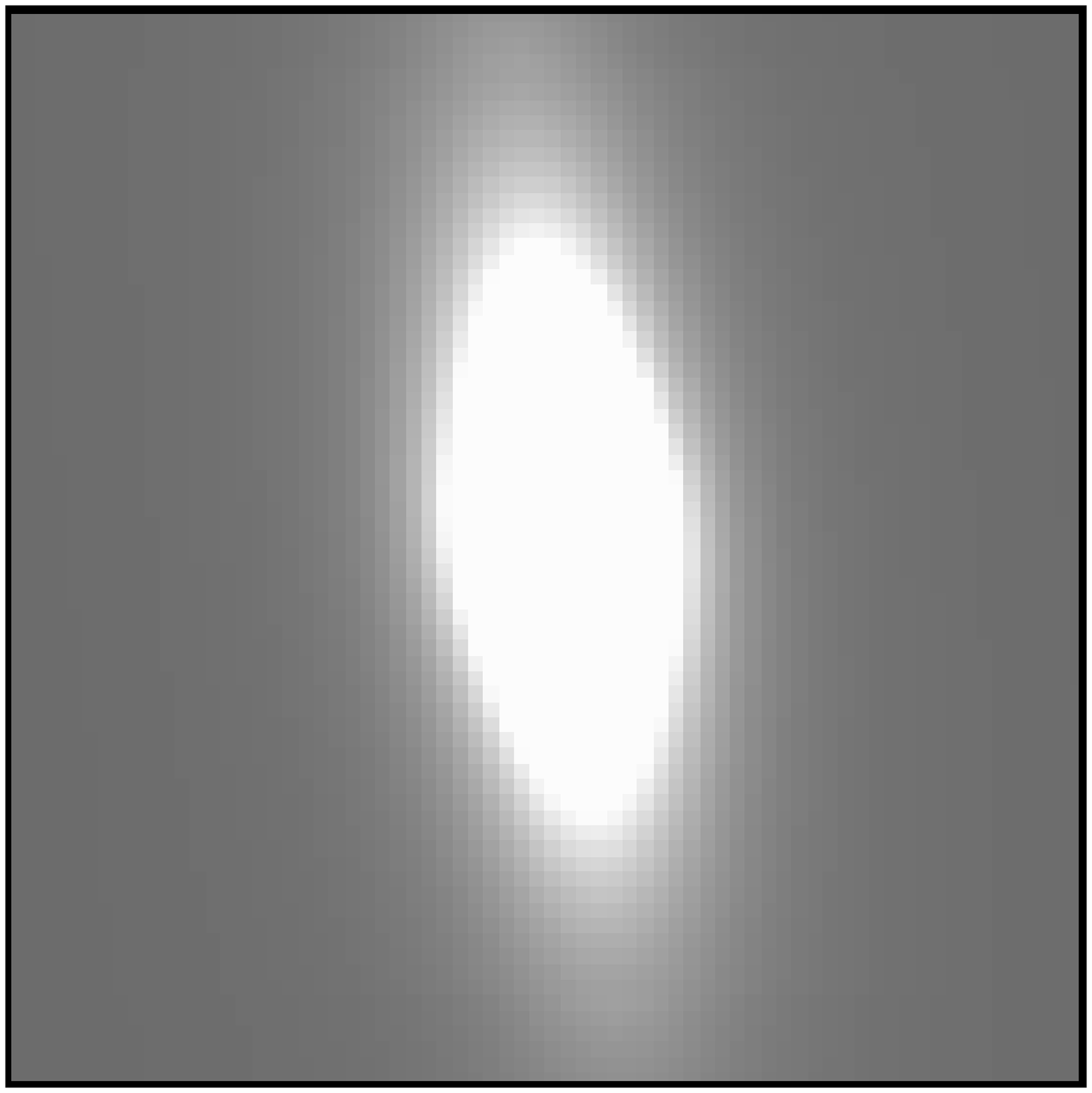}
\plotone{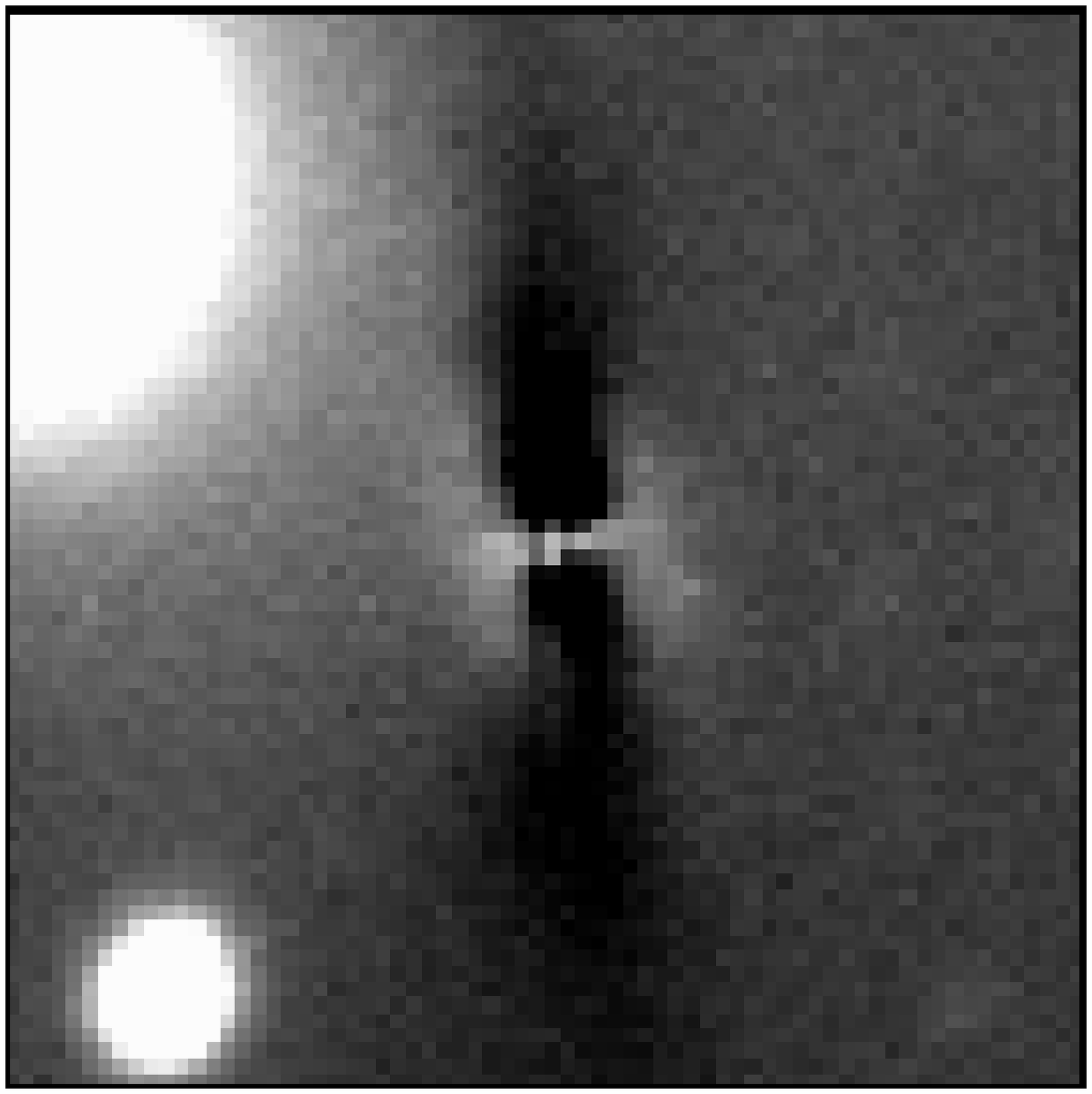}
\plotone{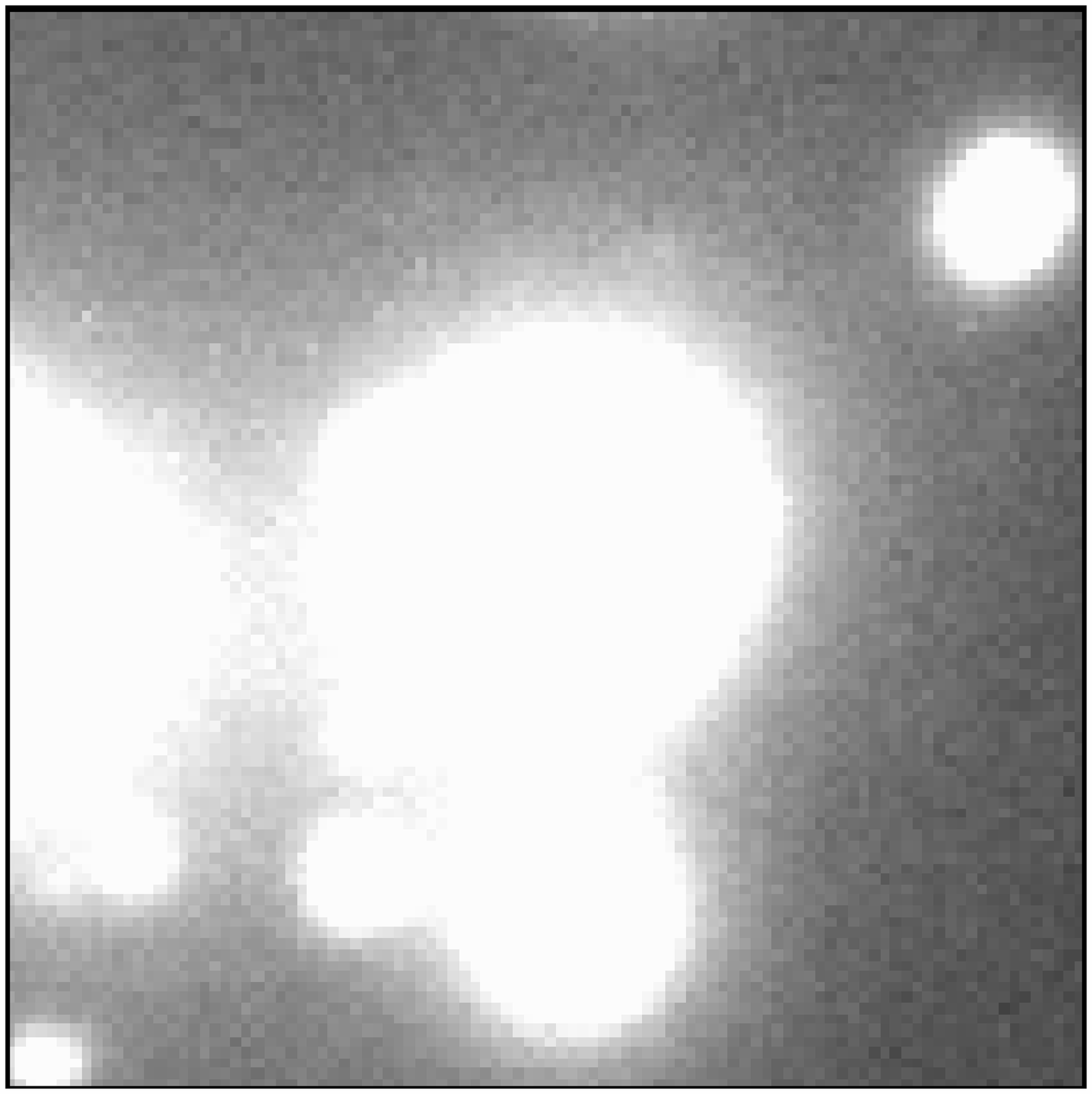}
\plotone{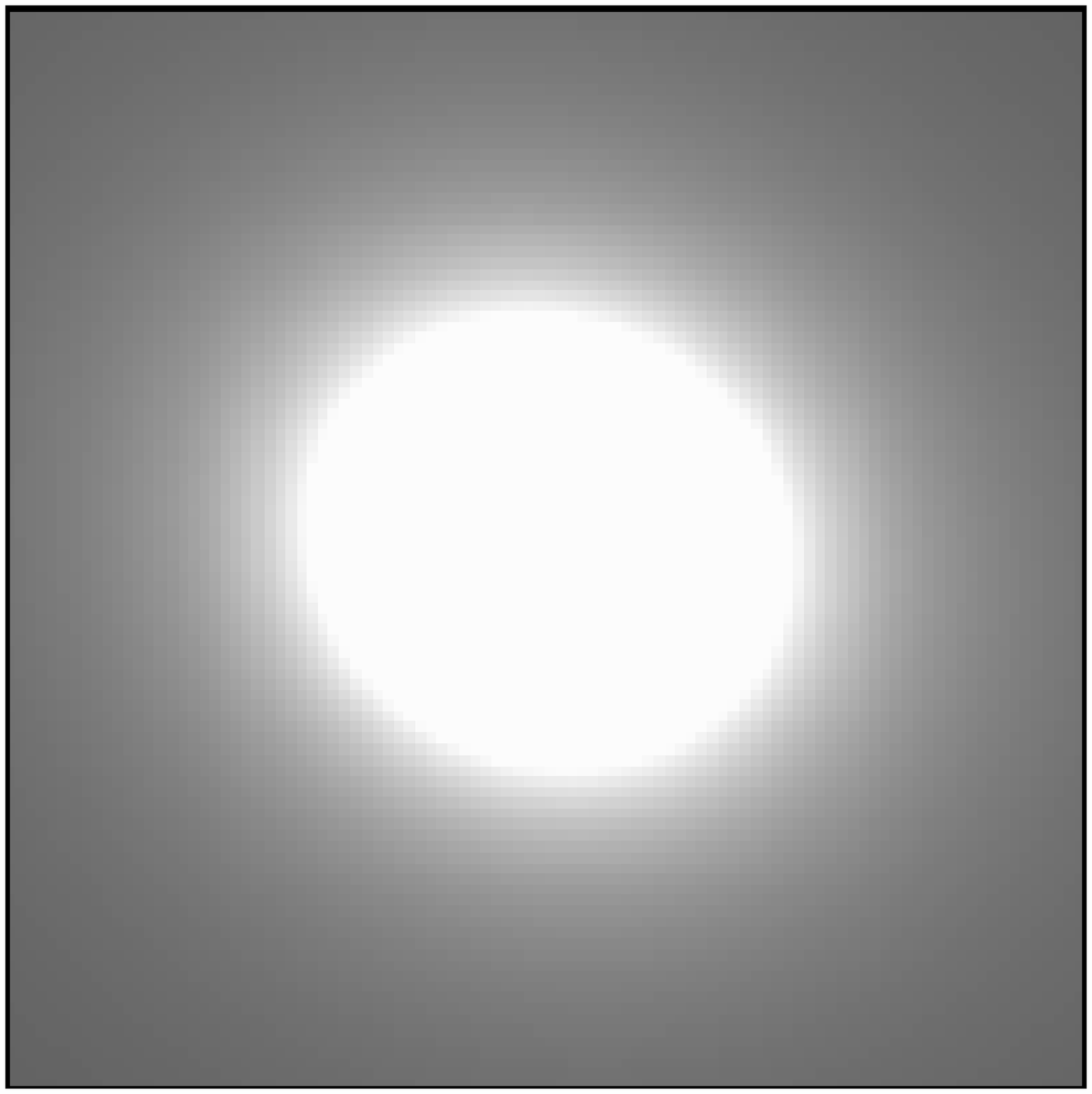}
\plotone{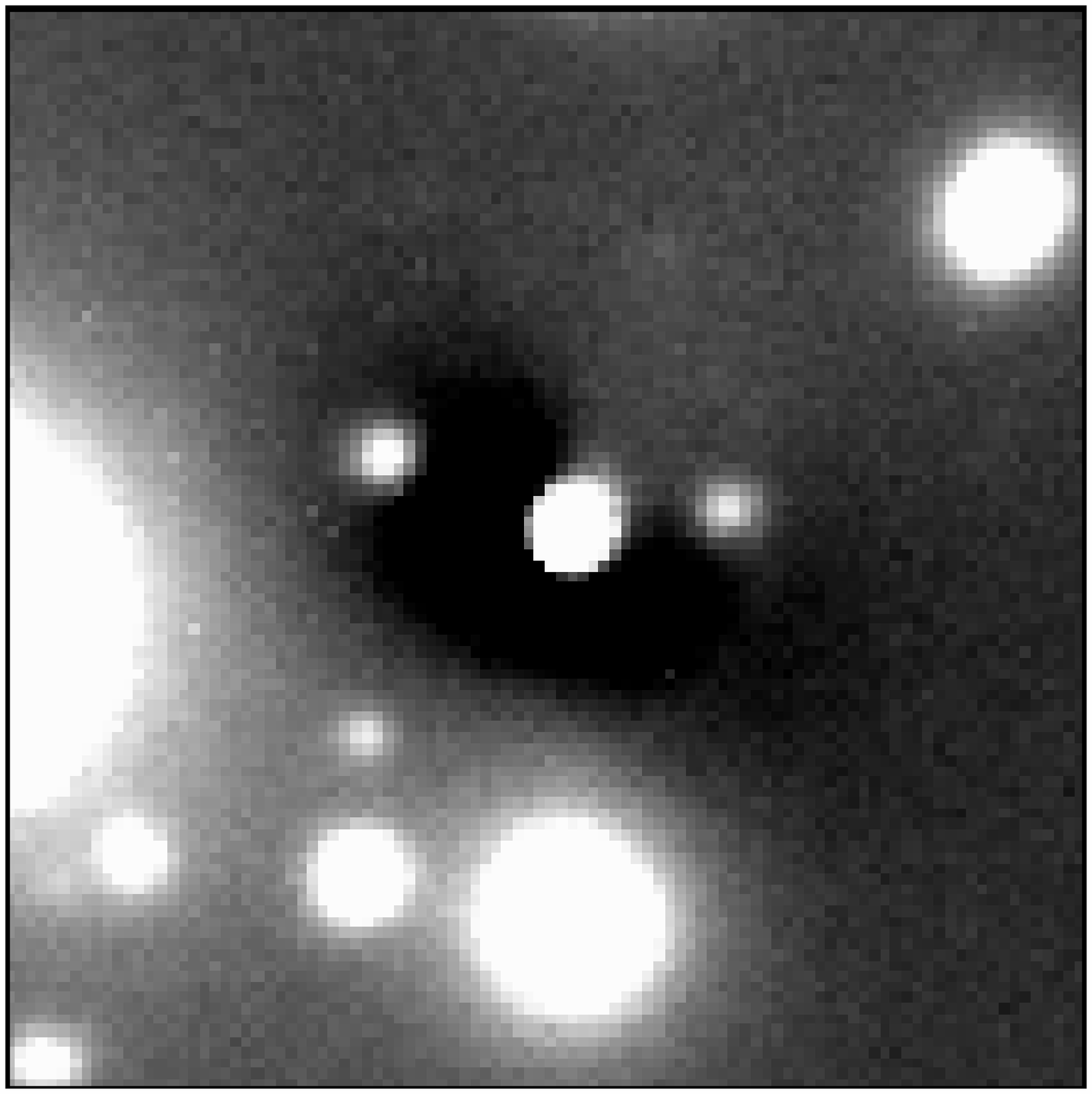}
\caption[galfit]{Images of observed galaxies, GALFIT models, and
  model-subtracted residual images are shown from left to right for 3
  different galaxies in A3888.  All images are shown at the same
  surface brightness levels.  The first row shows a fairly isolated
  galaxy in the outer regions of the cluster, which is well modeled by
  Galfit.  The second and third rows show galaxies in increasingly
  denser environments, depicting well the limitations of galaxy
  modeling algorithms for galaxies in very dense regions.}
\label{fig:galmodels}
\epsscale{1}
\end{figure}

Figure \ref{fig:galmodels} shows representative results of modeling 3
galaxies using Galfit: one isolated galaxy and two galaxies in
increasingly dense regions. These examples show that the algorithms
perform well for isolated galaxies, but fail for galaxies near the
core due either to difficulty in deblending many overlapping galaxy
profiles or because the individual galaxies in such dense regions do
not follow simple analytical profiles.  It is not clear what the
profiles should be of galaxies deep in the potential wells of clusters
\citep{trujillo2001a, feldmeier2004}.  The fact that A3888 is not a
relaxed cluster clearly makes galaxy subtraction more difficult near
the core than it would be in a CD cluster; A3888 has 3 main brightness
peaks which contain 3, 7, and 12 galaxy cores in their densest
regions, respectively.

As it is not possible to cleanly fit the galaxies in this cluster such
that the residuals (positive or negative) do not interfere with the
ICL measurement, we have chosen to mask the galaxies. This gives us a
well defined measurement of the ICL at the expense of forfeiting some
area. Although we could model and subtract the more isolated galaxies
in the outer regions of the cluster, it is in these regions that we
can generously mask the galaxies and still have enough pixels for an
ICL measurement.  Note that we do not replace masked pixels. Masked
regions are simply removed entirely from the ICL measurement.

We use the same masks for both bands so that all galaxies are masked
to the same radius, thereby insuring a self-consistent measurement of
the ICL color.  We use the $r-$band image to define the masks as it
has a deeper detection threshold (and thus larger detection areas)
than the $V-$band catalog.  Objects are identified using SExtractor
and masks are based on the isophotal detection area with a threshold
of 26.4 \magarc ($1.5\sigma$ above sky).  To be conservative in
rejection, we scale the semi-major and semi-minor axes identified by
SExtractor to increase the area of each galaxy mask by a
multiplicative factor of $2-2.3$, depending on the magnitude of the
galaxy.  To explore the effect of mask size on the profile shape of
the ICL, we make two additional images with mask sizes that are 30\%
smaller and 30\% larger than the original masks.  We then measure the
ICL three times with the three versions of mask sizes.  Additional
minor masking is done by hand to remove any remaining flux associated
with resolved objects.  These few regions are associated with small
overlapping sources which are not correctly deblended by SExtractor.

\begin{figure}
\plotone{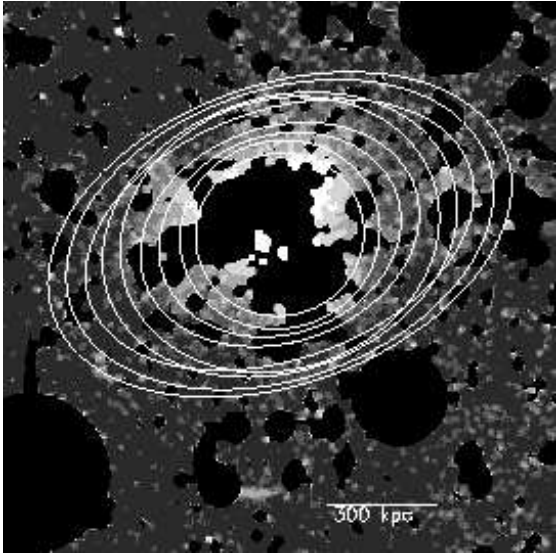}
\caption[isophotes]{ The fully masked, final $V-$band image of the central
  1.5 h$_{70}^{-1}$ Mpc (7.3 arcmin) of A3888, smoothed to aid in visual
  identification of surface brightness levels.  Masks are shown in
  their intermediate size, see \S \ref{galaxies}.  Large circular masks
  correspond to the locations of bright stars.  The six gray-scale
  levels show surface brightness levels of up to 28.5, 27.7, 27.2,
  26.7 \magarc, and brighter than 26.2\magarc. {\sc ellipse} isophotes
  are overlaid from 65 - 190\arcsec.  The tidal feature, C, shown also
  in Figure \ref{fig:arcs}, is clearly visible at center near the bottom
  of the image.}
\label{fig:isophotes}
\end{figure}

The total masked area within the central 1.2 Mpc of the cluster in
each of the three mask sizes is 34\%, 41\%, and 49\%.  The masked
fraction is much higher in the very center of the cluster and reaches
nearly 100\% in the inner 30 arcseconds.  The increase in masked
fraction is not directly proportional to the increase in mask size
because the masks often overlap. Figure \ref{fig:isophotes} shows the
final $V-$band image with intermediate-sized masks.

\subsection{Cluster Membership \& Flux}
\label{member}

An interesting characteristic of the ICL lies in its comparison to
cluster properties including the cluster galaxies themselves.  We
compare two methods below for measuring cluster membership and flux:
(1) we identify member galaxies using our own 2-band photometry; and
(2) we integrate the flux in a published galaxy luminosity function
for this cluster.

\begin{figure}
\plotone{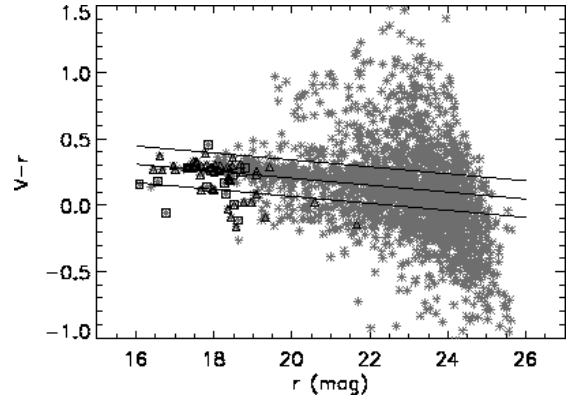}
\caption[cmd]{The color magnitude diagram of galaxies in A3888.  All
  galaxies detected in our data are plotted with gray symbols.  Those
  galaxies which have membership information in the literature are
  over-plotted with open triangles (members) or squares
  (non-members). The red sequence is clearly visible.  Solid lines
  indicate a biweight fit to the red sequence with $1\sigma$
  uncertainties.}
\label{fig:cmd}
\end{figure}

Some published velocities are available in the literature
\citep{teague1990, pimbblet2002} and can be used to explicitly
identify member galaxies. However these redshift surveys are not
complete to our detection threshold, and can therefore not provide
membership information for all detected galaxies.  Alternatively, we
can estimate cluster membership using a color magnitude relation
(Figure \ref{fig:cmd}) from our $V$ and gunn-$r$ images.  There is a
clear red sequence of galaxies where the brightest galaxies have
$V-r=0.3\pm0.15$.  Those galaxies which lie within 1$\sigma$ of a
biweight fit to the red sequence are taken to be cluster members
\citep[functional form taken from][]{beers1990}.  The slope of the red
sequence is 0.1 mags(color)/ 4mags(galaxy r magnitude).  Those
galaxies which are redder than the red sequence are both generally
fainter implying that they are higher redshift background galaxies and
are not as concentrated toward the center of the cluster as all
galaxies.  The number of those very red galaxies per projected area is
$38\pm 11\%$ higher within 400kpc than without. Although some of these
galaxies are undoubtedly members of the cluster, their spatial
distribution does not allow us to make conclusive statements about
their membership.  Approximately 42\% of the galaxies in the image are
identified as members by this method.  Of the galaxies with
spectroscopically determined velocities, 78\% of the 55 confirmed
members are included in the cut; 54\% of the 13 known non-members are
also included.  The red cluster sequence is a good tool for
identifying clusters, but it is not a perfect method of determining
membership as it is unable to cleanly distinguish between member and
non-member galaxies.

We measure the total flux in all galaxies identified as members
using corrected isophotal magnitudes from SExtractor. For these,
SExtractor assumes a Gaussian profile to infer the flux beyond the
isophotal detection threshold, corresponding to 26.0 $V$ \magarc\ and
26.4 $r$ \magarc.  As expected, the corrected magnitudes are brighter
than the isophotal magnitudes by a full magnitude at the faint end of
our detection limit.  The total flux in galaxies within
700h$_{70}^{-1}$ kpc center of the cluster, as determined from the
same galaxy catalog that was used for galaxy masking, is 3.9 x
$10^{12}$ \hbox{L$_{\odot}$} in the $V-$band and 4.9 x
$10^{12}$\hbox{L$_{\odot}$} in the $r-$band.  We expect the error on
the total flux from this estimate to be greater than 30\% which is
mainly due to uncertainty in the membership determination.

We can also determine cluster flux using the \citet{driver1998}
luminosity distribution for this cluster, which is based on a
statistical background subtraction of non-cluster galaxies.  It would
be possible to do this with our own data, however \citet{driver1998}
have more uniform, large-area coverage to several magnitudes below
$M^*$ at the redshift of the cluster.  In addition, the authors pay
careful attention to observing background fields which are up to
75\arcmin\ from the cluster center, at approximately the same airmass,
seeing, exptime, and UT as the cluster fields.  Consequently, the
background fields have the same noise characteristics and detection
threshold as the cluster images, and sample the same large scale
structures.  They can therefore be used to reliably determine
contamination of the cluster fields.  \citet{bernstein1995} gives a
careful account of the significant considerations in using this
method, all of which are taken into account by \citet{driver1998}.

We explore one minor effect not discussed by \citet{driver1998}: the
effect of gravitational lensing on the background galaxy counts. There
are two competing effects which change the number and brightness of
galaxies behind the cluster as compared to background galaxy counts in
an off-cluster field. First, magnification of the background
galaxies will artificially inflate the background counts behind
the cluster, resulting in an underestimation of cluster galaxy flux.
Second, all background objects behind the cluster will appear radially
more distant from the cluster center, which will artificially decrease
the background counts, resulting in an overestimate of the cluster
galaxy flux.  The change from an overall magnification to
de-magnification happens at $z \simeq 0.5$.  Following the method of
\citet{broadhurst1995} to determine the strength of the
de-magnification for A3888 at z = 0.15, we find a negligible
degradation in the $V-$ and $r-$ band flux ($< 0.2\%$), and therefore
do not correct for it in the \citet{driver1998} background counts.

\citet{driver1998} use their $R$-band luminosity distribution to
determine a dwarf to giant ratio, however we choose to fit it with a
classical Schechter function ($M^*_R = -22.82 \pm0.28, \alpha =
-0.97\pm0.09, \chi^2_{\nu}$ = 0.71), which can then be used to
determine a luminosity density for the cluster.  We note that the
luminosity distribution is not perfectly fit by a Schechter function
at the bright end, due mainly to a small number of extremely bright
galaxies, as is typical of clusters.  Adopting a volume equal to that
over which we are able to measure the ICL, 1.4 Mpc$^3$, and
integrating the luminosity function down to very faint dwarf galaxies,
$M_R$ = -11, the total luminosity from galaxies in the cluster is
$5.9\pm0.94\times 10^{12}$ \hbox{L$_{\odot}$} in the $R-$band.  Given
galaxy colors from \citet{fukugita1995}, the total luminosity from
galaxies in A3888 is 3.4 $\pm0.6\times 10^{12}$\hbox{L$_{\odot}$} in
$V$ and 4.3 $\pm0.7\times 10^{12}$\hbox{L$_{\odot}$} in the $r-$band.
The difference between this value of total flux and that determined
from our color--magnitude estimate of membership is likely due to
uncertainties in our membership identification and difference in
detection thresholds of the two surveys. Although the two estimates
are generally consistent, we adopt the total flux as derived from the
luminosity distribution throughout the remainder of the paper.


\section{Results}
\label{results}


\subsection{Surface brightness profile}
\label{profile}

After subtracting the stars and masking the galaxies, we fit the
resulting image with the IRAF routine {\sc ellipse}, a 2D,
interactive, isophote fitting algorithm.  Again, the masked pixels are
completely excluded in this procedure.  There are 3 free parameters in
the isophote fitting: center, position angle, and ellipticity.  We fix
the center ( J2000.0, $22^h 34^m 26.0^s, -37\degr 44\arcmin\
07.2\arcsec$) and position angle (-70 degrees) to values found by {\sc
ellipse} based on the inner isophotes, and let the ellipticity vary as
a function of radius.  Fitted ellipticities range from 0.2 to
0.5. Allowing the center and position angle to vary results in worse
fits.  Stable fits are found from $60-250\arcsec$.  From the fitted
isophotes we identify a fairly smooth ICL profile over the range of 26
to approximately 29 \magarc.  The error on the mean within each
elliptical isophote is negligible, as discussed in \S \ref{noise}.  It
is possible that the different seeing in the $V$ and $r-$band images
could unevenly affect the profiles.  To address this issue, the $V$
and $r-$band images have been convolved to the same seeing, and the
surface brightness profiles re-measured.  No significant change was
found in the profiles.

Note that we are not able to measure the ICL at radii smaller than
$60\arcsec$ because that region is heavily masked. Most other ICL
measurements focus on this inner region, leaving little overlap
between this survey and previous work in other clusters.  In clusters
containing a cD galaxy, the diffuse component of the cluster has been
found to blend smoothly into the cD envelope, and masking in the core
of such clusters is not necessary \citep[see most recently][]
{gonzalez2005}.

We identify the surface brightness profile of the total cluster light
(ie., including resolved galaxies) for comparison with the ICL within
the same radial extent. To do this, we make a new ``cluster'' image,
with color-determined, non-member galaxies masked out (see
\S\ref{member}).  A surface brightness profile of the cluster light is
then measured from this image using the same elliptical isophotes as
were used in the ICL profile measurement.  This profile, in contrast
to the ICL, is quite irregular, reflecting the clustering of galaxies.
Substructure in the galaxy distribution is an indication of a young
dynamical age for this cluster.

\begin{figure}
\plotone{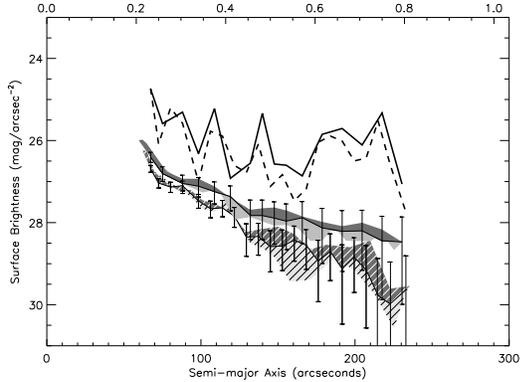}
\caption[icl]{ The surface brightness profile of the intracluster
  light as well as the total cluster light plotted as a function of
  distance along the semi-major axis in arcseconds.  The axis at the
  top of the figure indicates corresponding physical scale in
  h$_{70}^{-1}$Mpc.  We plot both $V-$ and $r-$ band data together for
  comparison.  The bottom two lines on the plot are the ICL profiles;
  the $r-$band light is surrounded by solid shading and the $V-$band
  is surrounded by hatched shading.  The shadings show the difference
  in ICL profiles produced by increasing or decreasing the area of the
  galaxy masks, as discussed in \S \ref{galaxies}.  The top two lines,
  without shading represent the total cluster light as measured in the
  same elliptical isophotes as the ICL; the dashed line represents the
  $V-$band light, the solid line represents the $r-$band.  Also shown
  are the cumulative $1\sigma$ errors for both bands as discussed in
  \S \ref{noise} and summarized in Table 3.} 
\label{fig:icl}
\end{figure}

Figure \ref{fig:icl} shows the surface brightness profiles of the ICL
as well as the total cluster light as a function of semi-major axis in
both the $V-$ and $r-$bands.  Results based on all three versions of
mask size (as discussed in \S \ref{galaxies}) are shown.  The
uncertainty in the ICL surface brightness is dominated by the accuracy
with which the background level can be identified, as discussed in \S
\ref{noise}.  Error bars in Figure \ref{fig:icl} show the cumulative
uncertainties tabulated in Table 3.

\begin{figure}
\plotone{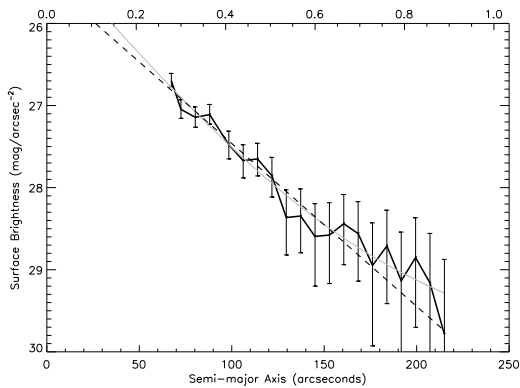}
\caption[Vexp]{The $V-$band intracluster
  light and $2\sigma$ error bars over-plotted with exponential fits.
  The best fit single(dashed) and double(gray) exponential are shown.}
\label{fig:iclVexp}
\end{figure}

\begin{figure}
\plotone{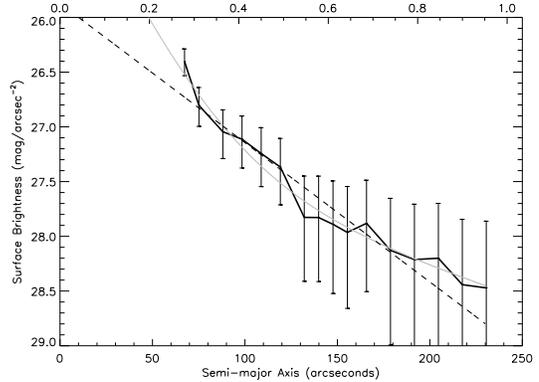}
\caption[rexp]{Same as Figure \ref{fig:iclVexp} in the $r-$band.}
\label{fig:iclrexp}
\end{figure}

Two characteristics are evident from the surface brightness profiles.
First, the inner region ($200-400$ h$_{70}^{-1}$kpc) has a notably
steeper profile than the outer region.  While the entire profile can
be adequately described within the $1\sigma$ uncertainties by a single
exponential, a double exponential gives a better fit in the $r-$ band
($\chi^2_{\nu}$ improves by $50\%$) and a marginally better fit in the
$V-$band.  These fits are shown in Figures \ref{fig:iclVexp} and
\ref{fig:iclrexp}.  We have also fit the ICL profile with
DeVaucouleurs and Sersic profiles.  Acceptable fits can be found,
however the best fit values are unphysical.  Namely they have high
exponents for the Sersic and unrealistically large effective radii for
the DeVaucouleurs profiles.  The second general characteristic of the
ICL is that it is more concentrated than the galaxies, which is to say
that the ICL falls off more rapidly with radius than the galaxy light.

\subsection{Spatial Distribution}

\begin{figure}
\plotone{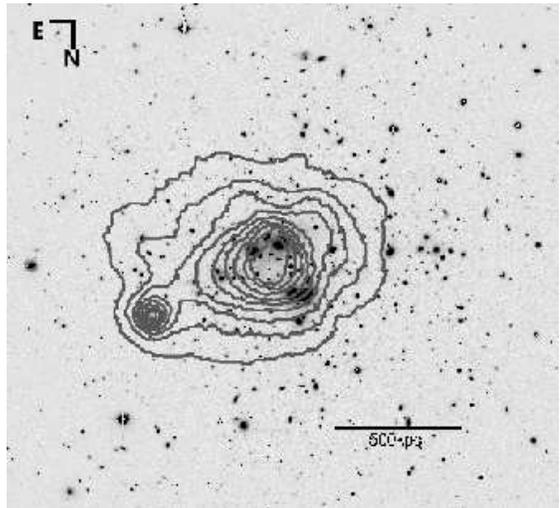}
\caption[X-ray contours]{ X-ray contours taken from XMM archival data
  are overlaid on our $V-$band optical image. Logarithmic contours are
  shown from 1 to 20 counts.  The bright point source 600h$_{70}^{-1}$ kpc from the
  cluster center is a Seyfert I galaxy.}
\label{fig:xray}
\end{figure}

The ICL is aligned to within $10^\circ$ of the position angle of the
hot intracluster gas.  Figure \ref{fig:xray} shows contours of XMM
archival observations overlaid on our optical image.  We interpret the
alignment of the diffuse intracluster light with the hot gas in the
cluster as an indication that we are indeed measuring light which
follows the gravitational potential of the cluster.  In addition, the
ICL radial surface brightness profile is significantly different than
the galaxy surface brightness profile in both $V$ and $r$, suggesting
that the intracluster light component is at least in part distinct
from the individual galaxies in the cluster.

\subsection{Color}

\begin{figure}
\plotone{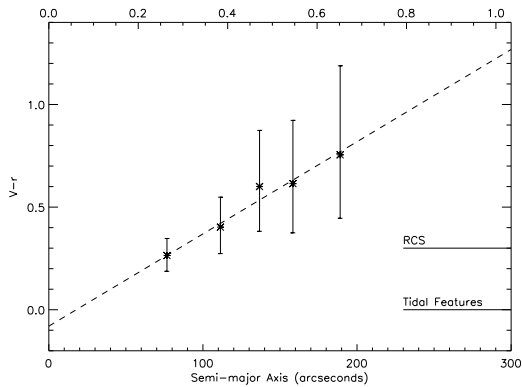}
\caption[color]{The ICL color versus radius in coarse radial bins
  based on the surface brightness in $V$ and $r$ as shown in Figure
  \ref{fig:icl}.  Lower axis shows radius in arcseconds, upper axis
  shows radius in Mpc.  The dashed line is the best fit linear
  function.  The average color of the red cluster sequence and the
  tidal features are also shown for comparison}
\label{fig:color}
\end{figure}

We measure an average $V-r$ color of the ICL by binning together three
to four points from the ICL radial profile.  Between 200 and
400h$_{70}^{-1}$ kpc, the innermost measured radii, the diffuse ICL
has an average color of $V-r \simeq 0.3\pm0.1$.  Beyond
400h$_{70}^{-1}$ kpc the ICL becomes increasingly redder, such that by
700h$_{70}^{-1}$ kpc the ICL has an average color of $V-r \simeq
0.7\pm 0.4$.  The only characteristic color of the galaxies we have to
compare with the ICL is the red sequence color ($V-r = 0.3\pm0.15$).
We have no definitive membership information for those galaxies off
the red sequence.  The color of the ICL in the inner 400h$_{70}^{-1}$
kpc is roughly equivalent to the red ellipticals residing in the same
part of the cluster, but significantly redder than several tidal
features we detect (see \S \ref{iclsub}).  The color of the ICL beyond
400h$_{70}^{-1}$ kpc is redder than the red sequence galaxies.  The
color of the diffuse ICL can be approximated as a simple linear
function of radius, with a slope of $+0.1$ per 100h$_{70}^{-1}$ kpc
and a y--intercept of $-0.1$.  Figure \ref{fig:color} shows the color
profile and corresponding $1\sigma$ error bars.  While this fit is
clearly simplistic, the data do not warrant a more complicated fit.
This red color gradient is opposite that which we expect to find for
the cluster galaxies.  When looking at the color of galaxies as a
function of distance from the center of the cluster, we find a flat or
slightly blueward profile such that the galaxies get slightly bluer
with increasing radius. Therefore the ICL color profile is distinct
from the galaxy color profile.

Using the population synthesis models of \citet{bruzual2003} we can
obtain rough constraints on the age and metallicity of the stellar
population which contributes to the ICL.  Because the total range in
color is not large and because the age-metallicity degeneracy limits
our conclusions, we limit our discussion to two regions,
$r<400$h$_{70}^{-1}$ kpc and $r>400$h$_{70}^{-1}$ kpc, rather than
individual points along the full radial profile.  The stellar
evolution models begin with a starburst of user defined strength and
extent, a Salpeter initial mass function, and a standard $\Lambda$CDM
Universe.  The stars then evolve along the Padova 1994 spectral
evolution tracks.  Within this scheme, the simplest scenario is an
instantaneous starburst with a single formation epoch and
metallicity. For this case, the red color of the ICL in the outer
$400-700 h_{70}^{-1}$kpc, $V-r \sim 0.6$, is consistent with a stellar
population which formed at redshifts $1 < $z$_{f}< 10$ (7 - 13 Gyr
ago) with an initial abundance of $1.0 Z_{\odot} < Z_{ICL} \la
2.5Z_{\odot}$.  The color of the ICL in the inner 200-400h$_{70}^{-1}$
kpc, $V-r \sim 0.3$, allows the minimum age of that range to be
lowered, where the most recent allowable formation is $\sim 5$ Gyr ago
($z_{f}< 1$) with an initial abundance of 0.2 - 0.5 solar.  Allowing
an extended burst of duration 10 to 100 Myr has a minimal effect on
the $V-r$ color.  Allowing an exponentially decaying star burst with
an e-folding time of 1Gyr, the population becomes overall 0.02 - 0.06
magnitudes bluer, depending on the initial formation redshift.  For
the ICL in A3888, an exponential star formation history therefore
implies even higher metallicities or earlier formation.  Finally,
simulations with a constant star formation rate of $1 \msun/$yr create
very blue stars.  It is not possible to form a stellar population with
a constant star formation rate which has $V-r = 0.6$.  Implications of
these models are discussed further in \S \ref{discusscolor}.


\subsection{Fractional Flux}

The ratio of ICL flux to total cluster flux can help constrain the
importance of galaxy disruption in the evolution of clusters.  To
identify the total flux in the ICL, we integrate the single
exponential fit to the ICL surface brightness profile (see
\S\ref{profile}) over the range 0--200\arcsec ($\sim 700$h$_{70}^{-1}$
kpc).  As we are not able to measure the ICL at radii smaller than
$\sim 60\arcsec$, this requires an extrapolation into the center of
the cluster.  Note that the single exponential fit, which is dominated
by the slope of the ICL profile at larger radii, gives significantly
less light in the core than the double exponential fit (see Figures
\ref{fig:iclVexp} and \ref{fig:iclrexp}), and is therefore a
conservative estimate of ICL flux.  The total flux in the
intracluster light is then $4.5\pm1.3\times 10^{11}$
\hbox{L$_{\odot}$} in $V$ and $5.9\pm2.2\times
10^{11}$\hbox{L$_{\odot}$} in $r$, where these errors are the full
errors as described in \S \ref{noise}.  This value is equivalent to
the full disruption of roughly $7 L^{*}$ galaxies.

We consider 4 modifications to this estimate of the total ICL flux.
First, we consider a correction for that volume of the cluster which
is filled with galaxies, since no ICL can exist in that volume.  While
lines of sight intersect galaxies over most of the area near the
center of the cluster, the galaxy filling factor is less than 3\% by
volume, even inside 200h$_{70}^{-1}$ kpc (60 arcsec, projected).  So
it is reasonable to assume that intracluster stars do exist in that
volume and we need make no correction for the filled volume.  Second,
we can determine a hard lower limit to the ICL flux by assuming that
there is {\it no} ICL in the inner 60\arcsec.  This correction, albeit
extreme, would decrease our estimate of the total ICL flux by 30\%.
Third, we make a less extreme correction by assuming a flat core
region instead of the exponential extrapolation.  A flat profile is
suggested by \citet{aguerri2005}, although those results are in Virgo,
where the center of the ICL is not defined and the measurement is
based on small area coverage, which does not allow for an elliptical
profile determination.  A flat core region would decrease our estimate
of the total ICL flux by 5\%.  Fourth, we consider low surface
brightness galaxies below our detection threshold which could
contribute to, and therefore be an error in, the inferred ICL flux.
To account for these very faint galaxies, we integrate the cluster
galaxy luminosity function from our detection limit ($M_R$ = -15.22)
to $M_R$ = -11.0.  Due to the extremely low detection threshold of
this survey (7.6 magnitudes below $M^*$), and the apparently flat
faint end of the luminosity function ($\alpha =-0.97$, see \S
\ref{member}), only 0.07\% of the total galaxy flux could come from
galaxies this faint.  As this contribution is not significant, we make
no correction for this effect.

Adopting the total galaxy flux found from the luminosity function in
\S \ref{member}, we find that the ICL accounts for $13\pm4\%$ of the
total $V-$band cluster light and $13\pm5\%$ of the $r-$band cluster
light within 700h$_{70}^{-1}$ kpc of the center of the cluster.  The
range in these values comes from the combination of all uncertainties
in the ICL measurement coupled with the uncertainty in the total
cluster flux, as discussed in \S \ref{noise}.  The galaxy light and
the ICL decrease with radius, but since we don't accurately know the
slope of either of them at large radii, we compare fluxes within the
same volume over which we have reliable data.  This fraction is only
relevant at this radius, and is likely to be lower when taking into
account the entire virial radius of the cluster, since the ICL is
centrally concentrated and not evenly distributed throughout the
cluster. On the same note, ICL measurements at smaller radii are
likely to find a higher fraction of the total flux in the ICL because
of the steep ICL profile, and because the volume involved is much
smaller.  For example, if we assume we can only measure the ICL in the
inner 600h$_{70}^{-1}$ instead of 700h$_{70}^{-1}$kpc, we find a
fractional flux of 19\% in both $V$ and $r$, an almost 50\% increase
over the measured 13\%.


\subsection{ICL Substructure}
\label{iclsub}

\begin{figure}
\includegraphics[angle=-90,scale=.20]{f9a.ps}
\includegraphics[angle=-90,scale=.20]{f9b.ps}
\plotone{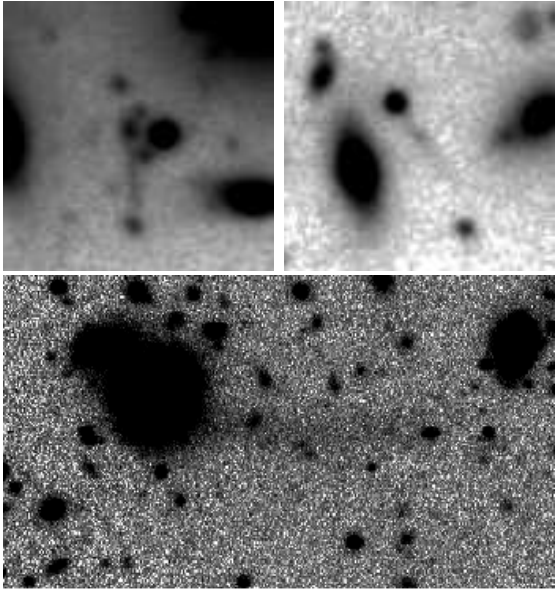}
\caption[arcs]{Three low surface brightness features from final
  $V-$band image.  The top two panels show features located near the
  center (features A \& B in Table 2).  The bottom
  image shows the larger feature (C) located $~500$h$_{70}^{-1}$ kpc from the center
  of the cluster.}
\label{fig:arcs}
\epsscale{1}
\end{figure}

Using the technique of unsharp masking, we find 3 possible tidal
features, all within the central $500$h$_{70}^{-1}$ kpc of A3888,
identified as A, B, and C in Figure \ref{fig:arcs}.  Arcs A and B are
both roughly $15$h$_{70}^{-1}$kpc $\times 5$h$_{70}^{-1}$kpc and are
near to the center of the cluster.  Arc C is a diffuse, tail-like
feature at $500$h$_{70}^{-1}$kpc from the center and covers
$130$h$_{70}^{-1}$kpc $\times 20$h$_{70}^{-1}$kpc (see Table 2).  All
three features are blue, $V-r \simeq 0.0$, with a combined flux
equivalent to one $r = 20.8$ magnitude galaxy (0.1$M^*$).  These
objects are unlikely to be gravitational arcs since they are not
oriented tangentially to the cluster potential.

The diffuse nature of the large feature, C, suggests that it is tidal.
This object is very similar to the large arcs found in Coma,
Centaurus, and A1914 \citep{gregg1998,calcaneo2000,feldmeier2004},
which are included in Table 2 for reference. In
general it is of similar size to, but slightly fainter than, those
found in the other clusters.  It has slight curvature and appears to
connect to a pair of galaxies (left side of the image as
displayed) that could be in the midst of an interaction.  Both
\citet{calcaneo2000} and \citet{feldmeier2004} find through numerical
simulations that these types of arcs are typical of recent tidal
interactions between luminous spiral galaxies and massive cluster
ellipticals.  Spectroscopy to confirm its origin at this faint surface
brightness is not currently possible.

\begin{deluxetable*}{lcccccccc}
\tabletypesize{\scriptsize}
\tablewidth{0pt}
\tablenum{2}
\tablecaption{Substructure in A3888
     \label{tab:features}}
\tablehead{
\colhead{Object} & \colhead{radial distance} & \colhead{length} & \colhead{width}& \colhead{$V-r$} & \colhead{M$_V$} & \colhead{M$_r$} & \colhead{$\mu(V)$}& \colhead{$\mu(r)$} \\
\colhead{} & \colhead{h$_{70}^{-1}$kpc} & \colhead{h$_{70}^{-1}$kpc} & \colhead{h$_{70}^{-1}$kpc} & \colhead{} & \colhead{} & \colhead{} & \colhead{mag arcsec$^{-2}$} & \colhead{mag arcsec$^{-2}$}
}

\startdata

A         & 66    & 17  & 5     & -0.05   &  -18.5  &  -18.5    & 24.6    & 24.5   \\
B         & 155   & 15  & 5     &  0      &  -18.1  &  -18.1    & 25.3    & 24.9   \\ 
C         & 720   & 132 & 20    & 0.03    &  -20.5  &  -20.5    & 26.4    & 26.7   \\ \\
\hline \\
Coma$^1$  & 100   & 130 & 15-30 & 0.57(R) &  -18.8  &  -19.4(R) & 26.9    & 26.3(R) \\
Centaurus$^2$ & 170   & 171 & 1     & \nodata      &  \nodata     &  -14.8(R) & \nodata      & 26.1(R)  \\
A1914$^3$     & 75    & 160 & 30    & \nodata      &  \nodata     &  \nodata       & 26.1    & \nodata     \\

\enddata

\tablecomments{
1: \citet{gregg1998}
2: \citet{calcaneo2000}
3: \citet{feldmeier2004}
}

\end{deluxetable*}

We cannot rule out the possibility that the 2 smaller features in our
0.06 degree$^{-2}$ of cluster imaging are low surface brightness (LSB)
galaxies seen edge-on.  In field surveys, surface densities of the
dimmest LSB galaxies ($23< \mu_0 < 25 V$ \magarc) are at least 0.01
galaxies degree$^{-2}$ \citep{dalcanton1997}.  In clusters, although
there are overall higher space densities of galaxies, LSB galaxies run
the risk of getting tidally disrupted in the harsh environment of
cluster centers.  In a survey of the Cancer and Pegasus clusters,
\citet{oneil1997} find 1.6 galaxies per square degree with central
surface brightness dimmer than 21.2 in $V$.  The 2 candidates with
average surface brightnesses of roughly 25\magarc\ in $V$ in this
cluster represent a higher density than found in these surveys.  In
addition, they do not have clear centers.  Both of these facts suggest
that they are not LSB galaxies.  However, it is likely that even if
these are LSB galaxies, they will not remain bound systems for long in
the high density cluster environment, and we therefore consider them
to be contributors to the ICL in the following calculation.

We briefly examine the importance of all three tidal features in
contributing to the ICL over a Hubble time to see if they can account,
in whole or in part, for the ICL found in the cluster.  Cluster
crossing time is estimated to be 4.5 Gyr given a virial radius of 3.7
Mpc and a temperature of 10KeV \citep{reiprich2002}.  We assume both a
constant rate of formation, and dissipation of tidal features in
approximately 1 crossing time.  From this we conclude that in 1 Hubble
time, approximately one half of an $M^*$ galaxy will be contributed to
the ICL through the visible tidal features such as these.  This simple
calculation suggests that these features cannot account for the
current ICL flux, however it is feasible that there was a variable
interaction rate in the history of this cluster.  Further substructure
could also be hidden below our surface brightness detection threshold.
  
At the distance of A3888, the flux of a single globular cluster ($M_V
\ga 10$ mag) spread over one seeing disk (3.5$h^{-1}_{70}$kpc)is many
magnitudes below our surface brightness detection threshold.
Therefore we are not sensitive to intracluster globular clusters which
have been studied by other groups in nearby clusters
\citep{jordan2003,bassino2003,hilker2003,marinfranch2003}.

\subsection{Group}

In addition to the main cluster ICL, we detect excess diffuse light
around a group of galaxies which are 1.35h$_{70}^{-1}$Mpc from the
center of A3888, (J2000) $22^h 34^m 48.5^s, -37\degr 39\arcmin\
19.58\arcsec$.  There are two galaxies centered in this diffuse
component, separated by only 2\arcsec.  The spatial extent of the
group appears to be 200h$_{70}^{-1}$ kpc, within which there are 60
galaxy peaks detected by SExtractor.  Independent of the ICL
component, the group is identified in the density distribution of
cluster.  Velocities are available only for the central galaxies in
the group, however these suggest that the group is co-spatial with
A3888 \citep{pimbblet2002, teague1990}.  Within the 200h$_{70}^{-1}$
kpc extent of the group, we find $1.7\times10^{10}$ \hbox{L$_{\odot}$}
in $V$ and $2.6\times 10^{10}$ \hbox{L$_{\odot}$} in $r$ above
background, which is equivalent to approximately $20\%$ of an $M^{*}$
galaxy.

The average color of this diffuse component based on total flux within
200h$_{70}^{-1}$ kpc is $V-r = 0.5$, which is again redder than the
cluster galaxies and consistent with the color of the main cluster ICL
at large radii.  The accuracy of the fluxes and hence the colors is
limited by the accuracy in masking since it is a simple sum over the
pixels in the group region.  We estimate the error in masking to be
less than 30\% based on our work with varying the mask size (see \S
\ref{profile} and Table 3).

This secondary ICL concentration is consistent with galaxy
interactions and ram pressure stripping occurring in an in-falling
group (``pre-processing'').  Such pre-processing has been shown in
simulations to affect galaxies before they fall into the main cluster
potential \citep[][and references therein]{willman2004,
fujita2004}. This is also consistent with recent measurements of a
small amount of ICL in isolated galaxy groups \citep{castro2003,
durrell2004}.

\newcommand\magarcs{mag arcsec$^{-2}$}

\begin{deluxetable*}{l c c c c c c c c}[h]
\tabletypesize{\scriptsize    }
\tablewidth{0pt}
\tablecaption{Error Budget
     \label{tab:error}}
\tablenum{3}
\tablehead{
\colhead{Source} & 
\colhead{} & 
\colhead{} & 
\multicolumn{6}{c}{contribution to ICL uncertainty (\%)} \\
\colhead{} & 
\multicolumn{2}{c}{$1\sigma$ uncertainty} &
\multicolumn{2}{c}{$\mu$(0\arcsec - 100\arcsec)} &
\multicolumn{2}{c}{$\mu$(100\arcsec - 200\arcsec)} &
\multicolumn{2}{c}{total ICL flux} \\
\colhead{} & 
\colhead{($V$)} & 
\colhead{($r$)} & 
\colhead{($V$)} & 
\colhead{($r$)} & 
\colhead{($V$)} & 
\colhead{($r$)} & 
\colhead{($V$)} & 
\colhead{($r$)}
}

\startdata

background level$^a$    & 29.5 \magarcs & 28.8 \magarcs & 14  & 18 & 39 & 45 & 24  & 31  \\
photometry              &  0.02 mag   & 0.03 mag    & 2   & 3  &  2 &  3 &  2  &  3  \\ 
masking$^b$             & \multicolumn{2}{c}{variation in mask area $\pm30$} & 5  & 5 & 14 & 19 & 9  & 12   \\
std.\ dev.\ in mean$^c$  &  32.7 \magarcs & 32.7 \magarcs & 3  & 2 & 2 & 1 & 3  & 1 \\ 
(total)                 & &                & 15 & 19 & 41 & 50 & 26 & 33\\ 
\\
\hline \\
cluster flux$^d$ &  16\% & 16\%  & \nodata & \nodata & \nodata & \nodata & \nodata  & \nodata  \\
\enddata

\tablecomments{ 
a: Large scale fluctuations in background level are measured
empirically and include instrumental calibration uncertainties 
as well as and true variations in background level (see \S \ref{noise}).
b: Object masks were scaled by $\pm30\%$ in area to test the impact on
ICL measurement (see \S\ref{galaxies}).
c: The statistical uncertainty in the mean surface brightness
of the ICL in each isophote.
d: Errors on the total cluster flux are based on errors in the fit to
the luminosity function (see \S \ref{member}). 
}

\end{deluxetable*}

\section{Accuracy Limits}
\label{noise}

\begin{figure}
\plotone{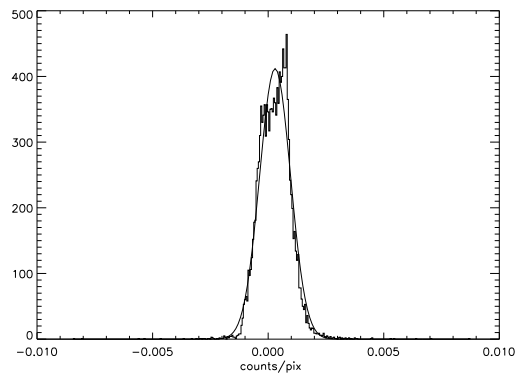}
\caption[Vexp]{Histogram of the mean values in $1\arcsec\ \times
  1\arcsec$ regions in counts in the fully masked $V-$ band image.
  All regions are greater than 800h$_{70}^{-1}$ kpc from the center of the cluster.
  A gaussian fit to the data is overlaid.}
\label{fig:noiseV}
\end{figure}

\begin{figure}
\plotone{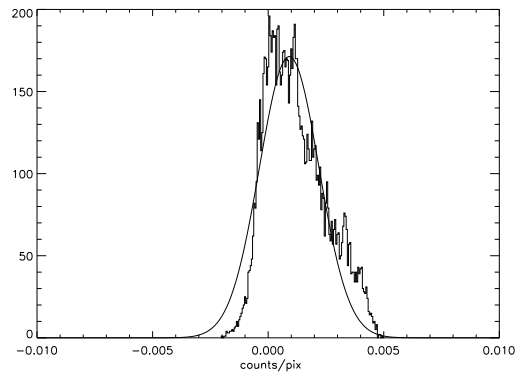}
\caption[Vexp]{Same as Figure \ref{fig:noiseV} in the $r-$band.}
\label{fig:noiser}
\end{figure}

The accuracy of the ICL surface brightness is limited on small scales
($<10\arcsec$) by photon noise.  On larger scales ($>10\arcsec$),
structure in the background level (be it physical on the sky or
instrumental) will dominate the error budget.  We determine the
stability of the background level in the image on large scales by
first median smoothing the masked image by 75\arcsec.  We then measure
the mean flux in thousands of random 1\arcsec\ regions more distant
than 0.8 Mpc from the center of the cluster.  The standard deviation
of these regions is 29.5 \magarc\ in $V$ (0.06\% of sky), and 28.8
\magarc\ in $r$ (0.01\% of sky).  Histograms with gaussians overlaid
are shown in Figures \ref{fig:noiseV} \& \ref{fig:noiser}.  The
histograms are not perfect gaussians.  This is likely due to the fact
that the background level includes both a symmetric gaussian and
positive sources which are below the detection threshold.  The offset
of the gaussian portion of the histogram represents the statistical
difficulty in measuring the mean value of the background in any one
image. Regions from all around the frame are used to check that our
accuracy limit is universal across the image and not affected by
location in the frame.  This empirical measurement of the large-scale
fluctuations across the image is dominated by the instrumental
flat-fielding accuracy, but includes contributions from the bias
and dark subtraction, physical variations in the sky level, and the
statistical uncertainties mentioned above.
 

This empirical measurement of the large--scale background fluctuations
is likely to be a worst--case estimate of the accuracy with which we
can measure surface brightness on large scales because it is derived
from the outer regions of the image where only 6-10 individual
exposures have been combined. In the central regions of our imaging
($r < 800$h$_{70}^{-1}$ kpc), roughly twice as many dithered images
have been combined, which has the effect of smoothing out large-scale
fluctuations in the illumination pattern to a greater degree.  We
therefore expect greater accuracy in the center of the image where the
ICL is being measured.

All sources of uncertainty are listed in Table 3.  In
addition to dominant uncertainty from our ability to measure the
large-scale fluctuations on the background as discussed above, we
quantify the contributions from the photometry, masking, and the
accuracy with which we can measure the mean in the individual
elliptical isophotes.  In total the error on the ICL flux is 26\% in V
and 33\% in r, which in addition to a 16\% error in the total cluster
flux, leads to a 30--40\% uncertainty in the fractional flux.

\section{Discussion}
\label{discuss}

We measure a diffuse intracluster component in A3888 to a radius of
$\sim 700$h$_{70}^{-1}$ kpc in the $V-$ and $r-$ band down to 28.9
\magarc\ and 28.2 \magarc\ respectively.  We discuss here the physical
implications of color, total flux, and profile shape of the ICL.


\subsection{Color Implications}
\label{discusscolor}

Color information can place constraints on the age and metallicity of
the progenitor population of the ICL, thereby shedding light on the
dominant physical mechanisms and timescales for galaxy disruption.
Color information may also be able to differentiate between the
morphological types of progenitor galaxies.

The color of the ICL in A3888 is consistent with some previous
observational results in other clusters, although those results vary
widely. \citet{schombert1988} and \citet{mackie1992} have found a wide
range of results for cD envelopes, from blue to red, with and without
color gradients. These surveys have typically been sensitive to a
diffuse component within 100-150h$_{70}^{-1}$ kpc, a much smaller
radial extent than this survey.  Recently, \citet{gonzalez2000} have
found a mild color gradient where the ICL becomes redder with radius
by $W-I = 0.25\pm0.08$ from 10 to 70$h^{-1}_{70}$kpc.  Again, our
observations cover a much larger radial region of A3888 and we have no
information on the ICL in the core region because it contains several
complicated, unmerged clumps.  Over a similar range in radius to our
measurement, \citet{zibetti2005}, from a stack of hundreds of SDSS
clusters, find an ICL including the BCG which is similar in color to
the galaxy light and has a flat or slightly blue color gradient with
radius.

If the ICL is composed of stars stripped from galaxies, its color
relative to the galaxies is indicative of the epoch when it was
stripped in the following sense. If the ICL is redder than the cluster
galaxies, it is likely to have been stripped from the galaxies at
early times (higher z).  Stripped stars will passively evolve toward
red colors, while the galaxies will continue to form stars. If, on the
other hand, the ICL is of similar color to the average galaxy, the ICL
is likely to have formed from the ongoing stripping of stars
\citep[via harassment as in ][]{moore1996}.  In this case the stripped
stars should have roughly the same color at the current epoch as the
galaxies at the current epoch.  This picture is complicated by the
fact that clusters are not made up of galaxies which were all formed
at a single epoch and that we don't know the star formation rates of
galaxies once they enter a cluster.  While these simple trends hold
for the colors of intracluster stars compared to galaxies, the color
difference between passively evolving stars and low star forming
galaxies may not be large enough to detect.

Cluster evolution is complex due to a myriad of environmental
influences. Several groups have produced hierarchical, $\Lambda$CDM
simulations of clusters which include radiative cooling, star
formation, and various feedback mechanisms, but differ primarily in
star formation prescriptions and numerical resolution.  These models
can be divided based on their broad, empirical predictions for the
color/formation epoch of the ICL.

Theoretical models in which the ICL forms early in the cluster history
all suggest an ICL which is older, redder than the galaxy population
because the galaxies continue to form new stars and therefore have
younger mean ages than the ICL population \citep{dubinski1998,
murante2004, sommer-larsen2005}.  This is generally consistent with
our results in the outer regions of A3888.  Specifically,
\citet{sommer-larsen2005} predict a slight color gradient in $B-R$
such that the ICL becomes 0.1 magnitudes bluer from 0 to
600h$_{70}^{-1}$ kpc, while our data suggest the opposite trend with
radius in $V-r$.

Theoretical models in which the ICL forms throughout the cluster
lifetime generally predict a younger, bluer intracluster stellar
population \citep{willman2004, moore1999, gnedin2003b, bekki2001}
since more recent stripping will have the chance to pull newly formed
stars out of galaxies.  Ongoing stripping is consistent with our
results within about 400h$_{70}^{-1}$ kpc, where the ICL is roughly
the same color or slightly bluer than the red cluster sequence.  In an
N-body + SPH simulation, \citet{willman2004} find that 50\% of
intracluster stars come from $M^{*}$ or brighter galaxies, which means
the color of the intracluster stars should be in accord with the color
of the outskirts of bright cluster galaxies or equivalently the color
of intermediate luminosity galaxies.  The intermediate luminosity
galaxies which we consider members of A3888 have a color in the range
of $V-r = 0.1 - 0.4$.  Our results are consistent with this prediction
in the inner regions of the cluster.

Recent observations of some intracluster HII regions \citep[such as
those found in Virgo by ][]{gerhard2002,ryan-weber2004,cortese2004}
indicate that it may be possible for some intracluster stars to form
in situ.  In this case, the ICL color will still depend on the
formation epoch.  If the ICSP is just now forming in intracluster HII
regions \citep[such as those found in Virgo by
][]{gerhard2002,ryan-weber2004,cortese2004} then it will be blue,
however if it formed earlier in cluster formation, then the ICSP will
passively evolve toward redder colors. Only if the ICL were
significantly bluer than the existing cluster galaxies could it be
possible to definitively state from the color that a significant
fraction of the ICL formed in situ.  Since this is not the case in
A3888, our results cannot constrain the formation site of the ICL.

Using N-body simulations, both \citet{moore1999} and
\citet{gnedin2003b} find that low density galaxies (LSB and dwarf
galaxies) are the main contributers to the ICL.  LSB galaxies in Sloan
data have a color range of $B-V = 0.4 - 1.3$ \citep{kniazev2004} which
corresponds to $V-r = -0.2 - 0.7$ .  Dwarf galaxies in Coma have $V-r
\simeq 0.5\pm0.3$ \citep{trentham1998}.  This range is sufficiently
broad that it is consistent with the ICL at all radii in A3888,
implying that the ICL could have origins in LSB or dwarf galaxies.

In summary, the ICL in the outer regions of A3888 is consistent
with the predictions for a stellar population which formed at
redshifts higher than 1 and is significantly metal-rich, implying an
ICL which forms early with the collapse of the main cluster.  The ICL
in the center of A3888 (r$ < 400$h$_{70}^{-1}$kpc) is consistent with
predictions for a relatively younger population.  This implies that
within some core radius harassment type interactions are the dominant
mechanism.  

\subsection{Fractional Flux Implications}

Another clue to the dominant mechanism driving evolution in clusters
comes from correlating ICL properties with the properties of the
parent cluster.  For example, a trend in ICL fraction with cluster
mass but not redshift, richness, or morphology would indicate that
mass was the dominant mechanism which could predict ICL fraction.  The
calculation of the fractional ICL flux depends on many observational
parameters including the surface brightness and radial limit of the
ICL measurement itself; the surface brightness at which individually
bound, resolved sources are distinguished from the ICL; and the volume
over which the ICL flux and galaxy flux are measured.  As these
parameters vary widely in work by previous groups, it is difficult to
make meaningful comparisons with results for other clusters in the
literature.  In addition, A3888 is a very massive cluster, and is not
as simple as clusters with a cD or clear brightest cluster galaxy
(BCG). A1914 is the only cluster with an ICL measurement
\citep{feldmeier2004} which has overall similar characteristics to
A3888.  With similar detection limits to those employed here, those
authors find an ICL fraction of 7\% in the $V-$band, which is
generally consistent with our results for A3888.

With these observational complications and cluster parameters in mind,
we can only generally conclude that previous measurements of the ICL
in clusters over a wide range in redshift ($0.003 < z < 0.41$) and
mass ($1 - 35 \times 10^{14}$\msun) are roughly 10\%.  There are no
obvious trends with mass or redshift, although there are some
noteworthy outliers at 50\% for Coma \citep{bernstein1995} and 0\% for
A1689 \citep{gudehus1989}.  The \citet{bernstein1995} result covers a
small radial extent and is therefore biased toward higher fractional
flux values.  It is difficult to interpret the \citet{gudehus1989}
measurement due to disparate methods.  There is a some evidence that
the ICL fraction is dependent upon cluster morphology; B/M type I
clusters \citep{theuns1997,feldmeier2002, uson1991,vg1994} have a
reported average ICL fraction which is marginally higher than those
with B/M type III \citep{vg1994,feldmeier2004, arnaboldi2003,
ferguson1998, durrell2002}.  However poor morphological
classification, small number statistics, and widely disparate methods
and accuracies among the different measurements make any possible
trends difficult to quantify.

In comparing the observed ICL with simulations, it is important to
note that the simulations generally report the fractional light in the
ICL out to much larger radii (e.g., $r_{200}$) than its surface
brightness can be measured observationally.  At smaller radii, the
predicted ratio of ICL to galaxy light would be larger.  Bearing this
in mind, \citet{willman2004} finds a lower limit for fractional flux
in the ICL of 10-22\% at $r_{200}$ for a Virgo-like cluster from
$z=1.1 - 0$ (increasing fractional flux with time). At the maximum
radius of our ICL measurement ($\sim0.3 r_{200}$), the fraction would
presumably be higher by at least a factor of two, making it larger
than we observe in A3888.  Other predictions are similarly high.  For
a cluster with the mass of A3888 ($25 \times 10^{14} h_{70}^{-1}$
\msun), both \citet{lin2004} and \citet{murante2004} predict an ICL
fraction in excess of 40\%.  To be consistent with their predictions,
this cluster would require a factor of greater than 100 lower mass to
have only 10\% ICL, and although this cluster is not dynamically
relaxed, such large erros in mass are not realistic.
 
In summary, we find an ICL fraction which is roughly compatible with
observed ICL measurements in other clusters.  However, our measurement
differs significantly from theoretical estimates, particularly
considering A3888's large total mass.  The dynamical state of A3888
may contribute to this discrepancy, as morphology may have a
significant influence over ICL fraction.  We emphasize again that
A3888 is not a relaxed cluster; it does not have a cD galaxy and it's
X-ray isophotes are not circular.  If we have caught this cluster as
it is just now entering its major merger phase, we would expect a low
ICL fraction as compared with a cluster at this redshift, mass, and
richness which had already reached dynamic equilibrium.  In contrast,
when examining Coma, a cluster with an extremely high ICL fraction but
lower mass, we note that it's morphology indicates that it has already
undergone significant merging to produce 2 cD-like galaxies.  If
morphology is the dominant influence on ICL flux, we should find A3888
to have a similar ICL fraction to other clusters with similar
morphologies.  Comparable measurements of the ICL in our remaining
sample will help to resolve this issue.

\subsection{Profile Shape Implications}

As discussed in \S\ref{profile}, the profile of the ICL is generally
stepper at smaller radii.  In particular, the steepening profile near
the core region of the cluster is associated with the 3 apparently
merging groups of galaxies in the center of the cluster.  The recent
interactions in the center have likely added and continue to add ICL,
which is likely to eventually relax into a BCG and BCG halo.  The
profile in the outer region of the cluster is consistent with previous
measurements of BCG envelopes which follow shallower profiles
\citep{gonzalez2005}.  In addition, \citet{bernstein1995} and
\citet{zibetti2005} find a steeper profile for the Coma cluster and
for a stacked profile of hundreds of Sloan clusters.  This steeper
profile is consistent with some theoretical predictions, particularly
by \citet{murante2004} based on a hydrodynamical simulation including
radiative cooling, star formation, and supernova feedback.



\section{Conclusion}
\label{conclude}

We have presented results for the first of ten clusters in our sample.
We have identified an intracluster component in A3888 to $\sim
700$h$_{70}^{-1}$ kpc from the center of the cluster down to $28.9$
\magarc\ in the $V-$ and $28.2$ \magarc\ in gunn-$r$ band.  This ICL
component is aligned with the hot gas in the cluster, which is
evidence of its correlation with the underlying mass distribution.
There is a second diffuse component around a group of galaxies
1.4h$_{70}^{-1}$Mpc from the center of the cluster which is consistent
with pre-processing in an in-falling group.  In addition to these two
diffuse ICL components, we find 3 low surface brightness features
consistent with being remnants from tidal interactions.

Beyond 400h$_{70}^{-1}$kpc from the center of the cluster, the ICL is
redder than the galaxies, implying an older population of stars.
Inside of 400h$_{70}^{-1}$kpc the ICL has a similar color to the
galaxies. We interpret this color gradient in the ICL ($V-r = 0.3 -
0.7$, from inner to outer) as evidence of younger intracluster stars
in the center of the cluster. Consequently, we suggest that more than
one process is likely stripping stars from cluster galaxies.
Specifically, harassment type interactions are still ongoing in the
center of the cluster while galaxy mergers may have played a
significant role earlier in the history of the cluster.

We find that the ICL component in A3888 does not follow the same light
profile as the resolved sources, but has a smoother and slightly
steeper profile than the galaxies. Due to a steepening profile within
400h$_{70}^{-1}$ kpc, the ICL profile can be described by a double
exponential function.  A double profile is consistent with ongoing
dynamical activity in the center of this cluster producing a new
population of intracluster stars.

Comparing the ICL to cluster galaxy flux, we find that the ICL
component in A3888 accounts for roughly $q13\%$ of the total cluster
flux within 700h$_{70}^{-1}$ kpc ($\sim 0.3 r_{200}$).  This value is
low compared to the theoretical predictions for a cluster of this
mass, and may be partly due to the fact that A3888 appears to be a
dynamically young cluster.  The ICL in A3888 will likely increase with
time due both to contributions from an in-falling group as well as
through major mergers in the center to create a cD galaxy.


\acknowledgments

We acknowledge C. Peng and L. Simard for help with their galaxy
profile fitting algorithms.  We thank the anonymous referee for useful
suggestions on the manuscript.  Partial support for J.E.K. was
provided by the National Science Foundation (NSF) through UM's NSF
ADVANCE program.  Partial support for R.A.B. was provided by a NASA
Hubble Fellowship grant HF-01088.01-97A awarded by Space Telescope
Science Institute, which is operated by the Association of
Universities for Research in Astronomy, Inc., for NASA under contract
NAS 5-2655.  This research has made use of data from the following
sources: USNOFS Image and Catalogue Archive operated by the United
States Naval Observatory, Flagstaff Station
(http://www.nofs.navy.mil/data/fchpix/); NASA/IPAC Extragalactic
Database (NED), which is operated by the Jet Propulsion Laboratory,
California Institute of Technology, under contract with the National
Aeronautics and Space Administration; the Two Micron All Sky Survey,
which is a joint project of the University of Massachusetts and the
Infrared Processing and Analysis Center/California Institute of
Technology, funded by the National Aeronautics and Space
Administration and the National Science Foundation; and the SIMBAD
database, operated at CDS, Strasbourg, France.


\bibliography{ms.2.bbl}  

\begin{thebibliography}{87}
\expandafter\ifx\csname natexlab\endcsname\relax\def\natexlab#1{#1}\fi

\bibitem[{{Abell} {et~al.}(1989){Abell}, {Corwin}, \& {Olowin}}]{abell1989}
{Abell}, G.~O., {Corwin}, H.~G., \& {Olowin}, R.~P. 1989, \apjs, 70, 1

\bibitem[{{Aguerri} {et~al.}(2005){Aguerri}, {Gerhard}, {Arnaboldi},
  {Napolitano}, {Castro-Rodriguez}, \& {Freeman}}]{aguerri2005}
{Aguerri}, J.~A.~L., {Gerhard}, O.~E., {Arnaboldi}, M., {Napolitano}, N.~R.,
  {Castro-Rodriguez}, N., \& {Freeman}, K.~C. 2005, \aj, 129, 2585

\bibitem[{Allen {et~al.}(2004)Allen, Schmidt, Ebeling, Fabian, \& van
  Speybroeck}]{allen2004}
Allen, S.~W., Schmidt, R.~W., Ebeling, H., Fabian, A.~C., \& van Speybroeck, L.
  2004, astro-ph/0405340

\bibitem[{{Arnaboldi} {et~al.}(2003){Arnaboldi}, {Freeman}, {Okamura},
  {Yasuda}, {Gerhard}, {Napolitano}, {Pannella}, {Ando}, {Doi}, {Furusawa},
  {Hamabe}, {Kimura}, {Kajino}, {Komiyama}, {Miyazaki}, {Nakata}, {Ouchi},
  {Sekiguchi}, {Shimasaku}, \& {Yagi}}]{arnaboldi2003}
{Arnaboldi}, M., {Freeman}, K.~C., {Okamura}, S., {Yasuda}, N., {Gerhard}, O.,
  {Napolitano}, N.~R., {Pannella}, M., {Ando}, H., {Doi}, M., {Furusawa}, H.,
  {Hamabe}, M., {Kimura}, M., {Kajino}, T., {Komiyama}, Y., {Miyazaki}, S.,
  {Nakata}, F., {Ouchi}, M., {Sekiguchi}, M., {Shimasaku}, K., \& {Yagi}, M.
  2003, \aj, 125, 514

\bibitem[{{Arnaboldi} {et~al.}(2004){Arnaboldi}, {Gerhard}, {Aguerri},
  {Freeman}, {Napolitano}, {Okamura}, \& {Yasuda}}]{arnaboldi2004}
{Arnaboldi}, M., {Gerhard}, O., {Aguerri}, J.~A.~L., {Freeman}, K.~C.,
  {Napolitano}, N.~R., {Okamura}, S., \& {Yasuda}, N. 2004, \apjl, 614, L33

\bibitem[{{Bassino} {et~al.}(2003){Bassino}, {Cellone}, {Forte}, \&
  {Dirsch}}]{bassino2003}
{Bassino}, L.~P., {Cellone}, S.~A., {Forte}, J.~C., \& {Dirsch}, B. 2003, \aap,
  399, 489

\bibitem[{{Batuski} {et~al.}(1999){Batuski}, {Miller}, {Slinglend},
  {Balkowski}, {Maurogordato}, {Cayatte}, {Felenbok}, \&
  {Olowin}}]{Batuski1999}
{Batuski}, D.~J., {Miller}, C.~J., {Slinglend}, K.~A., {Balkowski}, C.,
  {Maurogordato}, S., {Cayatte}, V., {Felenbok}, P., \& {Olowin}, R. 1999,
  \apj, 520, 491

\bibitem[{{Beers} {et~al.}(1990){Beers}, {Flynn}, \& {Gebhardt}}]{beers1990}
{Beers}, T.~C., {Flynn}, K., \& {Gebhardt}, K. 1990, \aj, 100, 32

\bibitem[{{Bekki} {et~al.}(2001){Bekki}, {Couch}, \& {Drinkwater}}]{bekki2001}
{Bekki}, K., {Couch}, W.~J., \& {Drinkwater}, M.~J. 2001, \apjl, 552, L105

\bibitem[{{Bernstein} {et~al.}(1995){Bernstein}, {Nichol}, {Tyson}, {Ulmer}, \&
  {Wittman}}]{bernstein1995}
{Bernstein}, G.~M., {Nichol}, R.~C., {Tyson}, J.~A., {Ulmer}, M.~P., \&
  {Wittman}, D. 1995, \aj, 110, 1507

\bibitem[{{Bertin} \& {Arnouts}(1996)}]{bertin1996}
{Bertin}, E. \& {Arnouts}, S. 1996, \aaps, 117, 393

\bibitem[{{Broadhurst} {et~al.}(1995){Broadhurst}, {Taylor}, \&
  {Peacock}}]{broadhurst1995}
{Broadhurst}, T.~J., {Taylor}, A.~N., \& {Peacock}, J.~A. 1995, \apj, 438, 49

\bibitem[{{Bruzual} \& {Charlot}(2003)}]{bruzual2003}
{Bruzual}, G. \& {Charlot}, S. 2003, \mnras, 344, 1000

\bibitem[{{Busarello} {et~al.}(2002){Busarello}, {Merluzzi}, {La Barbera},
  {Massarotti}, \& {Capaccioli}}]{Busarello2002}
{Busarello}, G., {Merluzzi}, P., {La Barbera}, F., {Massarotti}, M., \&
  {Capaccioli}, M. 2002, \aap, 389, 787

\bibitem[{{Calc{\' a}neo-Rold{\' a}n} {et~al.}(2000){Calc{\' a}neo-Rold{\'
  a}n}, {Moore}, {Bland-Hawthorn}, {Malin}, \& {Sadler}}]{calcaneo2000}
{Calc{\' a}neo-Rold{\' a}n}, C., {Moore}, B., {Bland-Hawthorn}, J., {Malin},
  D., \& {Sadler}, E.~M. 2000, \mnras, 314, 324

\bibitem[{{Castro-Rodr{\'{\i}}guez} {et~al.}(2003){Castro-Rodr{\'{\i}}guez},
  {Aguerri}, {Arnaboldi}, {Gerhard}, {Freeman}, {Napolitano}, \&
  {Capaccioli}}]{castro2003}
{Castro-Rodr{\'{\i}}guez}, N., {Aguerri}, J.~A.~L., {Arnaboldi}, M., {Gerhard},
  O., {Freeman}, K.~C., {Napolitano}, N.~R., \& {Capaccioli}, M. 2003, \aap,
  405, 803

\bibitem[{{Chen} {et~al.}(1998){Chen}, {Huchra}, {McNamara}, \&
  {Mader}}]{Chen1998}
{Chen}, J., {Huchra}, J.~P., {McNamara}, B.~R., \& {Mader}, J. 1998, Bulletin
  of the American Astronomical Society, 30, 1307

\bibitem[{{Ciardullo} {et~al.}(1985){Ciardullo}, {Ford}, \&
  {Harms}}]{Ciardullo1985}
{Ciardullo}, R., {Ford}, H., \& {Harms}, R. 1985, \apj, 293, 69

\bibitem[{{Collins} {et~al.}(1995){Collins}, {Guzzo}, {Nichol}, \&
  {Lumsden}}]{Collins1995}
{Collins}, C.~A., {Guzzo}, L., {Nichol}, R.~C., \& {Lumsden}, S.~L. 1995,
  \mnras, 274, 1071

\bibitem[{{Cortese} {et~al.}(2004){Cortese}, {Gavazzi}, {Boselli}, \&
  {Iglesias-Paramo}}]{cortese2004}
{Cortese}, L., {Gavazzi}, G., {Boselli}, A., \& {Iglesias-Paramo}, J. 2004,
  \aap, 416, 119

\bibitem[{{Couch} {et~al.}(2001){Couch}, {Balogh}, {Bower}, {Smail},
  {Glazebrook}, \& {Taylor}}]{Couch2001}
{Couch}, W.~J., {Balogh}, M.~L., {Bower}, R.~G., {Smail}, I., {Glazebrook}, K.,
  \& {Taylor}, M. 2001, \apj, 549, 820

\bibitem[{{Couch} \& {Newell}(1984)}]{Couch1984}
{Couch}, W.~J. \& {Newell}, E.~B. 1984, \apjs, 56, 143

\bibitem[{{Couch} \& {Sharples}(1987)}]{Couch1987}
{Couch}, W.~J. \& {Sharples}, R.~M. 1987, \mnras, 229, 423

\bibitem[{{Dalcanton} {et~al.}(1997){Dalcanton}, {Spergel}, {Gunn}, {Schmidt},
  \& {Schneider}}]{dalcanton1997}
{Dalcanton}, J.~J., {Spergel}, D.~N., {Gunn}, J.~E., {Schmidt}, M., \&
  {Schneider}, D.~P. 1997, \aj, 114, 635

\bibitem[{{De Filippis} {et~al.}(2004){De Filippis}, {Bautz}, {Sereno}, \&
  {Garmire}}]{filippis2004}
{De Filippis}, E., {Bautz}, M.~W., {Sereno}, M., \& {Garmire}, G.~P. 2004,
  \apj, 611, 164

\bibitem[{{De Propris} {et~al.}(2002){De Propris}, {Couch}, {Colless},
  {Dalton}, {Collins}, {Baugh}, {Bland-Hawthorn}, {Bridges}, {Cannon}, {Cole},
  {Cross}, {Deeley}, {Driver}, {Efstathiou}, {Ellis}, {Frenk}, {Glazebrook},
  {Jackson}, {Lahav}, {Lewis}, {Lumsden}, {Maddox}, {Madgwick}, {Moody},
  {Norberg}, {Peacock}, {Percival}, {Peterson}, {Sutherland}, \&
  {Taylor}}]{DePropris2002}
{De Propris}, R., {Couch}, W.~J., {Colless}, M., {Dalton}, G.~B., {Collins},
  C., {Baugh}, C.~M., {Bland-Hawthorn}, J., {Bridges}, T., {Cannon}, R.,
  {Cole}, S., {Cross}, N., {Deeley}, K., {Driver}, S.~P., {Efstathiou}, G.,
  {Ellis}, R.~S., {Frenk}, C.~S., {Glazebrook}, K., {Jackson}, C., {Lahav}, O.,
  {Lewis}, I., {Lumsden}, S., {Maddox}, S., {Madgwick}, D., {Moody}, S.,
  {Norberg}, P., {Peacock}, J.~A., {Percival}, W., {Peterson}, B.~A.,
  {Sutherland}, W., \& {Taylor}, K. 2002, \mnras, 329, 87

\bibitem[{{den Hartog}(1995)}]{denHartog1995}
{den Hartog}, R. 1995, Ph.D.~Thesis

\bibitem[{{Domainko} {et~al.}(2004){Domainko}, {Gitti}, {Schindler}, \&
  {Kapferer}}]{domainko2004}
{Domainko}, W., {Gitti}, M., {Schindler}, S., \& {Kapferer}, W. 2004, \aap,
  425, L21

\bibitem[{{Driver} {et~al.}(1998){Driver}, {Couch}, \&
  {Phillipps}}]{driver1998}
{Driver}, S.~P., {Couch}, W.~J., \& {Phillipps}, S. 1998, \mnras, 301, 369

\bibitem[{{Dubinski}(1998)}]{dubinski1998}
{Dubinski}, J. 1998, \apj, 502, 141

\bibitem[{{Durrell} {et~al.}(2002){Durrell}, {Ciardullo}, {Feldmeier},
  {Jacoby}, \& {Sigurdsson}}]{durrell2002}
{Durrell}, P.~R., {Ciardullo}, R., {Feldmeier}, J.~J., {Jacoby}, G.~H., \&
  {Sigurdsson}, S. 2002, \apj, 570, 119

\bibitem[{{Durrell} {et~al.}(2004){Durrell}, {Decesar}, {Ciardullo},
  {Hurley-Keller}, \& {Feldmeier}}]{durrell2004}
{Durrell}, P.~R., {Decesar}, M.~E., {Ciardullo}, R., {Hurley-Keller}, D., \&
  {Feldmeier}, J.~J. 2004, in IAU Symposium, 90

\bibitem[{{Ebeling} {et~al.}(1996){Ebeling}, {Voges}, {Bohringer}, {Edge},
  {Huchra}, \& {Briel}}]{ebeling1996}
{Ebeling}, H., {Voges}, W., {Bohringer}, H., {Edge}, A.~C., {Huchra}, J.~P., \&
  {Briel}, U.~G. 1996, \mnras, 281, 799

\bibitem[{{Feldmeier} {et~al.}(2004){Feldmeier}, {Mihos}, {Morrison},
  {Harding}, {Kaib}, \& {Dubinski}}]{feldmeier2004}
{Feldmeier}, J.~J., {Mihos}, J.~C., {Morrison}, H.~L., {Harding}, {Kaib},
  P.~N., \& {Dubinski}, J. 2004, \apj, in press

\bibitem[{{Feldmeier} {et~al.}(2002){Feldmeier}, {Mihos}, {Morrison}, {Rodney},
  \& {Harding}}]{feldmeier2002}
{Feldmeier}, J.~J., {Mihos}, J.~C., {Morrison}, H.~L., {Rodney}, S.~A., \&
  {Harding}, P. 2002, \apj, 575, 779

\bibitem[{{Ferguson} {et~al.}(1998){Ferguson}, {Tanvir}, \& {von
  Hippel}}]{ferguson1998}
{Ferguson}, H.~C., {Tanvir}, N.~R., \& {von Hippel}, T. 1998, \nat, 391, 461

\bibitem[{{Fujita}(2004)}]{fujita2004}
{Fujita}, Y. 2004, \pasj, 56, 29

\bibitem[{{Fukugita} {et~al.}(1995){Fukugita}, {Shimasaku}, \&
  {Ichikawa}}]{fukugita1995}
{Fukugita}, M., {Shimasaku}, K., \& {Ichikawa}, T. 1995, \pasp, 107, 945

\bibitem[{{Gerhard} {et~al.}(2005){Gerhard}, {Arnaboldi}, {Freeman},
  {Kashikawa}, {Okamura}, \& {Yasuda}}]{gerhard2005}
{Gerhard}, O., {Arnaboldi}, M., {Freeman}, K.~C., {Kashikawa}, N., {Okamura},
  S., \& {Yasuda}, N. 2005, \apjl, 621, L93

\bibitem[{{Gerhard} {et~al.}(2002){Gerhard}, {Arnaboldi}, {Freeman}, \&
  {Okamura}}]{gerhard2002}
{Gerhard}, O., {Arnaboldi}, M., {Freeman}, K.~C., \& {Okamura}, S. 2002, \apjl,
  580, L121

\bibitem[{{Girardi} {et~al.}(1998){Girardi}, {Borgani}, {Giuricin},
  {Mardirossian}, \& {Mezzetti}}]{Girardi1998b}
{Girardi}, M., {Borgani}, S., {Giuricin}, G., {Mardirossian}, F., \&
  {Mezzetti}, M. 1998, \apj, 506, 45

\bibitem[{{Girardi} {et~al.}(1997){Girardi}, {Escalera}, {Fadda}, {Giuricin},
  {Mardirossian}, \& {Mezzetti}}]{Girardi1997}
{Girardi}, M., {Escalera}, E., {Fadda}, D., {Giuricin}, G., {Mardirossian}, F.,
  \& {Mezzetti}, M. 1997, \apj, 482, 41

\bibitem[{{Girardi} \& {Mezzetti}(2001)}]{girardi2001}
{Girardi}, M. \& {Mezzetti}, M. 2001, \apj, 548, 79

\bibitem[{{Gnedin}(2003)}]{gnedin2003b}
{Gnedin}, O.~Y. 2003, \apj, 589, 752

\bibitem[{{Gonzalez} {et~al.}(2005){Gonzalez}, {Zabludoff}, \&
  {Zaritsky}}]{gonzalez2005}
{Gonzalez}, A.~H., {Zabludoff}, A.~I., \& {Zaritsky}, D. 2005, \apj, 618, 195

\bibitem[{{Gonzalez} {et~al.}(2000){Gonzalez}, {Zabludoff}, {Zaritsky}, \&
  {Dalcanton}}]{gonzalez2000}
{Gonzalez}, A.~H., {Zabludoff}, A.~I., {Zaritsky}, D., \& {Dalcanton}, J.~J.
  2000, \apj, 536, 561

\bibitem[{{Gregg} \& {West}(1998)}]{gregg1998}
{Gregg}, M.~D. \& {West}, M.~J. 1998, \nat, 396, 549

\bibitem[{{Gudehus}(1989)}]{gudehus1989}
{Gudehus}, D.~H. 1989, \apj, 340, 661

\bibitem[{{Hilker}(2003)}]{hilker2003}
{Hilker}, M. 2003, in Extragalactic Globular Cluster Systems, 173

\bibitem[{{Jedrzejewski}(1987)}]{Jedrzejewski1987}
{Jedrzejewski}, R.~I. 1987, \mnras, 226, 747

\bibitem[{{Jord{\' a}n} {et~al.}(2003){Jord{\' a}n}, {West}, {C{\^ o}t{\' e}},
  \& {Marzke}}]{jordan2003}
{Jord{\' a}n}, A., {West}, M.~J., {C{\^ o}t{\' e}}, P., \& {Marzke}, R.~O.
  2003, \aj, 125, 1642

\bibitem[{{Jorgensen}(1994)}]{jorgensen1994}
{Jorgensen}, I. 1994, \pasp, 106, 967

\bibitem[{{Kelson} {et~al.}(2002){Kelson}, {Zabludoff}, {Williams}, {Trager},
  {Mulchaey}, \& {Bolte}}]{kelson2002}
{Kelson}, D.~D., {Zabludoff}, A.~I., {Williams}, K.~A., {Trager}, S.~C.,
  {Mulchaey}, J.~S., \& {Bolte}, M. 2002, \apj, 576, 720

\bibitem[{{Kniazev} {et~al.}(2004){Kniazev}, {Grebel}, {Pustilnik}, {Pramskij},
  {Kniazeva}, {Prada}, \& {Harbeck}}]{kniazev2004}
{Kniazev}, A.~Y., {Grebel}, E.~K., {Pustilnik}, S.~A., {Pramskij}, A.~G.,
  {Kniazeva}, T.~F., {Prada}, F., \& {Harbeck}, D. 2004, \aj, 127, 704

\bibitem[{{Kowalski} {et~al.}(1983){Kowalski}, {Ulmer}, \&
  {Cruddace}}]{Kowalski1983}
{Kowalski}, M.~P., {Ulmer}, M.~P., \& {Cruddace}, R.~G. 1983, \apj, 268, 540

\bibitem[{{Landolt}(1992)}]{landolt1992}
{Landolt}, A.~U. 1992, \aj, 104, 340

\bibitem[{{Leinert} {et~al.}(1998){Leinert}, {Bowyer}, {Haikala}, {Hanner},
  {Hauser}, {Levasseur-Regourd}, {Mann}, {Mattila}, {Reach}, {Schlosser},
  {Staude}, {Toller}, {Weiland}, {Weinberg}, \& {Witt}}]{leinert1998}
{Leinert}, C., {Bowyer}, S., {Haikala}, L.~K., {Hanner}, M.~S., {Hauser},
  M.~G., {Levasseur-Regourd}, A.-C., {Mann}, I., {Mattila}, K., {Reach}, W.~T.,
  {Schlosser}, W., {Staude}, H.~J., {Toller}, G.~N., {Weiland}, J.~L.,
  {Weinberg}, J.~L., \& {Witt}, A.~N. 1998, \aaps, 127, 1

\bibitem[{{Lin} \& {Mohr}(2004)}]{lin2004}
{Lin}, Y. \& {Mohr}, J.~J. 2004, \apj, 617, 879

\bibitem[{{Mackie}(1992)}]{mackie1992}
{Mackie}, G. 1992, \apj, 400, 65

\bibitem[{{Mar{\'{\i}}n-Franch} \& {Aparicio}(2003)}]{marinfranch2003}
{Mar{\'{\i}}n-Franch}, A. \& {Aparicio}, A. 2003, \apj, 585, 714

\bibitem[{{Markevitch}(1998)}]{markevitch1998}
{Markevitch}, M. 1998, \apj, 504, 27

\bibitem[{{Mazure} {et~al.}(1996){Mazure}, {Katgert}, {den Hartog}, {Biviano},
  {Dubath}, {Escalera}, {Focardi}, {Gerbal}, {Giuricin}, {Jones}, {Le Fevre},
  {Moles}, {Perea}, \& {Rhee}}]{Mazure1996}
{Mazure}, A., {Katgert}, P., {den Hartog}, R., {Biviano}, A., {Dubath}, P.,
  {Escalera}, E., {Focardi}, P., {Gerbal}, D., {Giuricin}, G., {Jones}, B., {Le
  Fevre}, O., {Moles}, M., {Perea}, J., \& {Rhee}, G. 1996, \aap, 310, 31

\bibitem[{{Moore} {et~al.}(1996){Moore}, {Katz}, {Lake}, {Dressler}, \&
  {Oemler}}]{moore1996}
{Moore}, B., {Katz}, N., {Lake}, G., {Dressler}, A., \& {Oemler}, A. 1996,
  \nat, 379, 613

\bibitem[{{Moore} {et~al.}(1999){Moore}, {Lake}, {Stadel}, \&
  {Quinn}}]{moore1999}
{Moore}, B., {Lake}, G., {Stadel}, J., \& {Quinn}, T. 1999, in ASP Conf. Ser.
  170: The Low Surface Brightness Universe, 229

\bibitem[{{Murante} {et~al.}(2004){Murante}, {Arnaboldi}, {Gerhard}, {Borgani},
  {Cheng}, {Diaferio}, {Dolag}, {Moscardini}, {Tormen}, {Tornatore}, \&
  {Tozzi}}]{murante2004}
{Murante}, G., {Arnaboldi}, M., {Gerhard}, O., {Borgani}, S., {Cheng}, L.~M.,
  {Diaferio}, A., {Dolag}, K., {Moscardini}, L., {Tormen}, G., {Tornatore}, L.,
  \& {Tozzi}, P. 2004, \apjl, 607, L83

\bibitem[{{O'Neil} {et~al.}(1997){O'Neil}, {Bothun}, \& {Cornell}}]{oneil1997}
{O'Neil}, K., {Bothun}, G.~D., \& {Cornell}, M.~E. 1997, \aj, 113, 1212

\bibitem[{{Peng} {et~al.}(2002){Peng}, {Ho}, {Impey}, \& {Rix}}]{peng2002}
{Peng}, C.~Y., {Ho}, L.~C., {Impey}, C.~D., \& {Rix}, H. 2002, \aj, 124, 266

\bibitem[{{Pimbblet} {et~al.}(2002){Pimbblet}, {Smail}, {Kodama}, {Couch},
  {Edge}, {Zabludoff}, \& {O'Hely}}]{pimbblet2002}
{Pimbblet}, K.~A., {Smail}, I., {Kodama}, T., {Couch}, W.~J., {Edge}, A.~C.,
  {Zabludoff}, A.~I., \& {O'Hely}, E. 2002, \mnras, 331, 333

\bibitem[{{Reimers} {et~al.}(1996){Reimers}, {Koehler}, \&
  {Wisotzki}}]{reimers1996}
{Reimers}, D., {Koehler}, T., \& {Wisotzki}, L. 1996, \aaps, 115, 235

\bibitem[{{Reiprich} \& {B{\" o}hringer}(2002)}]{reiprich2002}
{Reiprich}, T.~H. \& {B{\" o}hringer}, H. 2002, \apj, 567, 716

\bibitem[{{Ryan-Weber} {et~al.}(2004){Ryan-Weber}, {Meurer}, {Freeman},
  {Putman}, {Webster}, {Drinkwater}, {Ferguson}, {Hanish}, {Heckman},
  {Kennicutt}, {Kilborn}, {Knezek}, {Koribalski}, {Meyer}, {Oey}, {Smith},
  {Staveley-Smith}, \& {Zwaan}}]{ryan-weber2004}
{Ryan-Weber}, E.~V., {Meurer}, G.~R., {Freeman}, K.~C., {Putman}, M.~E.,
  {Webster}, R.~L., {Drinkwater}, M.~J., {Ferguson}, H.~C., {Hanish}, D.,
  {Heckman}, T.~M., {Kennicutt}, R.~C., {Kilborn}, V.~A., {Knezek}, P.~M.,
  {Koribalski}, B.~S., {Meyer}, M.~J., {Oey}, M.~S., {Smith}, R.~C.,
  {Staveley-Smith}, L., \& {Zwaan}, M.~A. 2004, \aj, 127, 1431

\bibitem[{{Schombert}(1988)}]{schombert1988}
{Schombert}, J.~M. 1988, \apj, 328, 475

\bibitem[{{Simard} {et~al.}(2002){Simard}, {Willmer}, {Vogt}, {Sarajedini},
  {Phillips}, {Weiner}, {Koo}, {Im}, {Illingworth}, \& {Faber}}]{Simard2002}
{Simard}, L., {Willmer}, C.~N.~A., {Vogt}, N.~P., {Sarajedini}, V.~L.,
  {Phillips}, A.~C., {Weiner}, B.~J., {Koo}, D.~C., {Im}, M., {Illingworth},
  G.~D., \& {Faber}, S.~M. 2002, \apjs, 142, 1

\bibitem[{{Sommer-Larsen} {et~al.}(2005){Sommer-Larsen}, {Romeo}, \&
  {Portinari}}]{sommer-larsen2005}
{Sommer-Larsen}, J., {Romeo}, A.~D., \& {Portinari}, L. 2005, \mnras, 39

\bibitem[{{Stein}(1996)}]{Stein1996}
{Stein}, P. 1996, \aaps, 116, 203

\bibitem[{{Struble} \& {Ftaclas}(1994)}]{struble1994}
{Struble}, M.~F. \& {Ftaclas}, C. 1994, \aj, 108, 1

\bibitem[{{Struble} \& {Rood}(1999)}]{Struble1999}
{Struble}, M.~F. \& {Rood}, H.~J. 1999, \apjs, 125, 35

\bibitem[{{Teague} {et~al.}(1990){Teague}, {Carter}, \& {Gray}}]{teague1990}
{Teague}, P.~F., {Carter}, D., \& {Gray}, P.~M. 1990, \apjs, 72, 715

\bibitem[{{Theuns} \& {Warren}(1997)}]{theuns1997}
{Theuns}, T. \& {Warren}, S.~J. 1997, \mnras, 284, L11

\bibitem[{{Trentham}(1998)}]{trentham1998}
{Trentham}, N. 1998, \mnras, 293, 71

\bibitem[{{Trujillo} {et~al.}(2001){Trujillo}, {Aguerri}, {Guti{\' e}rrez}, \&
  {Cepa}}]{trujillo2001a}
{Trujillo}, I., {Aguerri}, J.~A.~L., {Guti{\' e}rrez}, C.~M., \& {Cepa}, J.
  2001, \aj, 122, 38

\bibitem[{{Uson} {et~al.}(1991){Uson}, {Boughn}, \& {Kuhn}}]{uson1991}
{Uson}, J.~M., {Boughn}, S.~P., \& {Kuhn}, J.~R. 1991, \apj, 369, 46

\bibitem[{{Vilchez-Gomez} {et~al.}(1994){Vilchez-Gomez}, {Pello}, \&
  {Sanahuja}}]{vg1994}
{Vilchez-Gomez}, R., {Pello}, R., \& {Sanahuja}, B. 1994, \aap, 283, 37

\bibitem[{{Willman} {et~al.}(2004){Willman}, {Governato}, {Wadsley}, \&
  {Quinn}}]{willman2004}
{Willman}, B., {Governato}, F., {Wadsley}, J., \& {Quinn}, T. 2004, \mnras,
  355, 159

\bibitem[{{Wu} {et~al.}(1999){Wu}, {Xue}, \& {Fang}}]{Wu1999}
{Wu}, X., {Xue}, Y., \& {Fang}, L. 1999, \apj, 524, 22

\bibitem[{{Zaritsky} {et~al.}(2004){Zaritsky}, {Gonzalez}, \&
  {Zabludoff}}]{zaritsky2004}
{Zaritsky}, D., {Gonzalez}, A.~H., \& {Zabludoff}, A.~I. 2004, \apjl, 613, L93

\bibitem[{{Zibetti} {et~al.}(2005){Zibetti}, {White}, {Schneider}, \&
  {Brinkmann}}]{zibetti2005}
{Zibetti}, S., {White}, S.~D.~M., {Schneider}, D.~P., \& {Brinkmann}, J. 2005,
  \mnras, 358, 949

\end{thebibliography}

\clearpage

\begin{figure}
\plotone{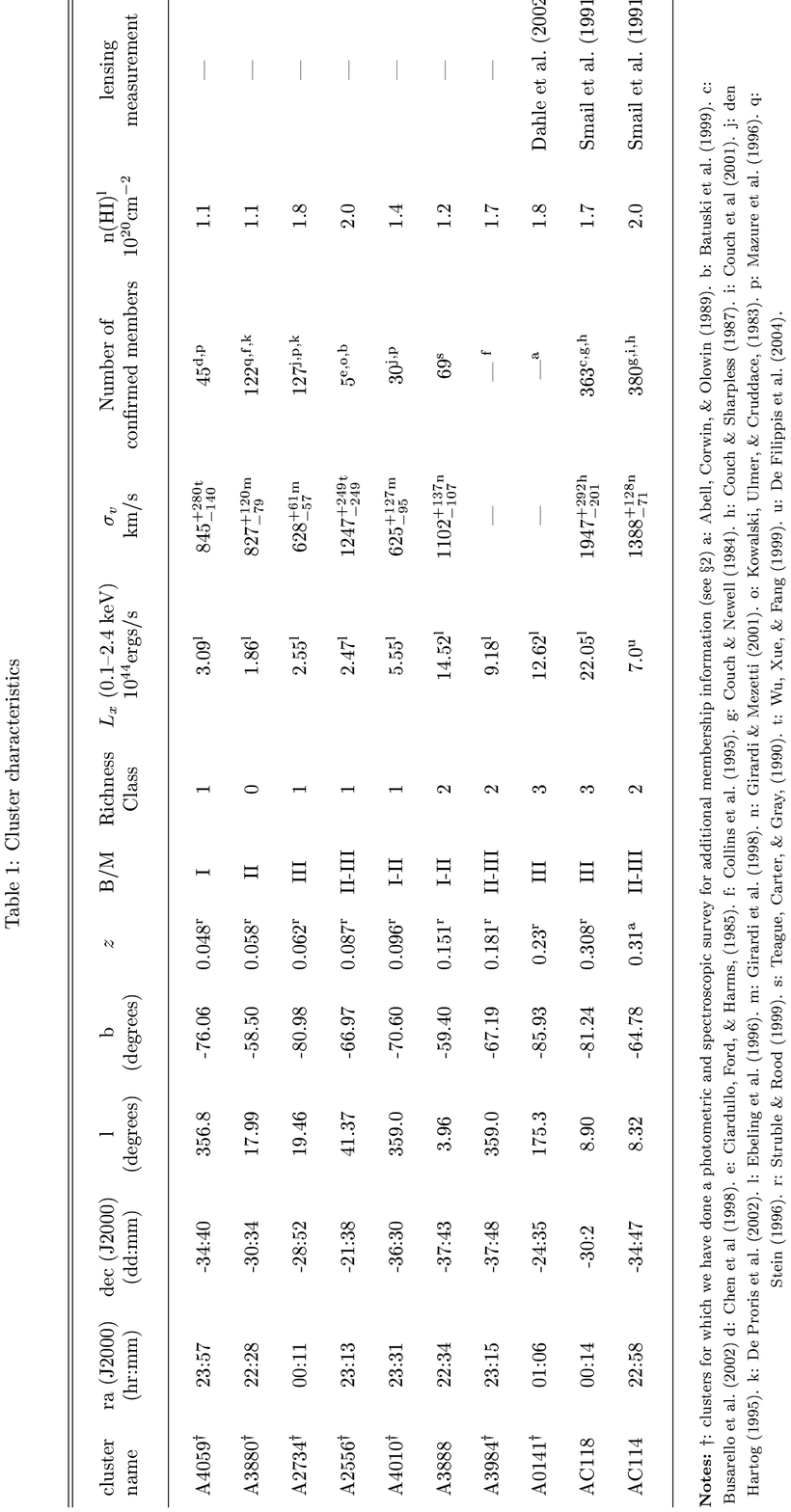}
\label{fig:table }
\end{figure}

\end{document}